\documentclass[aps,pre,superscriptaddress,twocolumn,floatfix]{revtex4-1}
\usepackage{bm}
\usepackage{bbm}
\usepackage[normalem]{ulem}
\usepackage{dcolumn}
\usepackage{color}
\usepackage{xcolor}
\usepackage{xfrac}
\usepackage{mathtools}
\usepackage{hyperref}
\usepackage{kotex}
\usepackage{graphicx} 
\usepackage{bm}
\usepackage{amsmath}
\usepackage{amssymb}
\usepackage{graphicx}
\usepackage{slashed}
\usepackage{mathrsfs}
\usepackage{mdframed}
 \usepackage{cancel}
 \usepackage{setspace}
  \usepackage{xcolor}
 \usepackage{babel}
 \usepackage{graphicx} 
 \usepackage{pifont}
 \usepackage{cancel} 
 \usepackage{hyperref} 
 \usepackage{listings} 
 \usepackage{verbatim} 
\usepackage{bbm}
\usepackage[normalem]{ulem}
\usepackage{amsmath}
\usepackage{graphicx}   
\usepackage{float}      

\begin{document}
\title{Role of System-Bath Interaction in Non-Markovian Quantum Brownian Otto Cycles}
\author{Haena Shim}
\affiliation{Department of Physics, Konkuk University, Seoul 05029, Korea}
\author{Joonhyun Yeo} 
\affiliation{Department of Physics, Konkuk University, Seoul 05029, Korea}
\date{\today}

\begin{abstract}

We study finite-time quantum Otto cycles whose working medium is a harmonic oscillator
undergoing a quantum Brownian motion described by the Caldeira-Leggett model
when the oscillator is in contact with heat baths
in isochoric processes. The time evolution of the Otto cycle is studied by
analytically solving the exact Heisenberg-Langevin equations for the system variables and 
the interaction energy between the system and the bath. This enables 
us to investigate non-Markovian strong-coupling effects on the quantum Otto cycle. 
We obtain cyclic steady states  and study the thermodynamic properties of the Otto cycle
for various values of the parameters describing the heat baths and the coupling between 
the system and the bath. We compare our results with those obtained in
the Markovian limit, where the time evolution is described by the Lindblad equation.
We find that the change in the interaction energy during the isochoric process 
contributes to both work and heat and plays a crucial role in determining the
thermodynamic behavior of the cycle. 
This is in contrast to the behavior of the state observables
of the working medium, which show relatively small change compared to the Markovian counterparts.
We find that non-Markovian effects only become significant when we consider thermodynamic
fluxes such as work and heat with the system-bath interaction fully taken into account.
We also compare our results with the power-efficiency trade-off relation 
recently proposed for the Markovian quantum Otto engine. 

\end{abstract}

\maketitle
\section{INTRODUCTION}

Quantum heat engines, as devices that can convert one form
of energy into another at the nanoscale, have been the subject of theoretical
\cite{Kosloff_2014,Cangemi_Bhadra_Levy_2024} and 
experimental \cite{Abah_2012,Ros_2016,Myers_Abah_Deffner_2022} interest. 
They are also important at a more fundamental level, as they can provide a useful
framework for testing and applying quantum thermodynamics \cite{Binder2019-hy,Kosloff_2013,Vinjanampathy_Anders_2016}, where thermodynamic concepts
such as heat, work and the Carnot bound are extended to small systems governed by quantum mechanics.

The study of quantum heat engines usually involves a quantum system of a working medium
interacting with an environment consisting of a heat bath and, therefore, requires methods developed in the theory of open quantum systems \cite{OQS_Breuer,rivas2012open,QDS_Weiss}.
Quantum Otto cycles have been studied extensively \cite{Feldmann_Kosloff_2004,Rezek_Kosloff_2006,Agarwal_Chaturvedi_2013,Zheng_Poletti_2014,Kosloff_Rezek_2017,Insinga_2018,Kloc_Cejnar_Schaller_2019,Abah_Paternostro_2019,Park_Lee_Chun_Noh_2019,Chen_Sun_Dong_2019,Dann_Kosloff_Salamon_2020,lee2020} mainly because they 
involve a time-independent working medium in contact with
the heat bath, which may allow analytic treatment of the dynamics.
The time evolution of such an open quantum system is usually described by 
quantum master equations. Conventionally, most studies on quantum Otto cycles have
used the Lindblad equation \cite{Lindblad,GKS} for the time evolution of the 
working medium, which is based on the Markovian approximation. 
For a general system plus reservoir setting, the derivation of 
the Lindblad equation involves the Born-Markov approximation followed by the 
secular approximation \cite{OQS_Breuer,rivas2012open}. The combination of these
approximations is valid in the limit where the coupling between the system and the reservoir is vanishingly small \cite{rivas2012open,trushechkin_open_2022}.

However, there are situations where the Markovian approximation is not applicable.
This is especially the case when the coupling between the system and the bath is not negligible. Recently, there has been a surge of interest in 
the study of non-Markovian effects on quantum Otto cycles 
\cite{Zhang_2014, Pozas-Kerstjens_Brown_Hovhannisyan_2018, Thomas_Siddharth_Banerjee_Ghosh_2018,Pezzutto_Paternostro_Omar_2019, Mukherjee_Kofman_Kurizki_2020, Wiedmann_2020,Liu_Jung_Segal_2021,Wiedmann_2021,Shirai_2021,Chakraborty_2022, Ptaszy_2022,Cavaliere_2022,Carrega_2022,Arisoy_2022,Razzoli_2023,Ishizaki_2023,Maity_Ghoshal_2024,Picatoste_2024}. An early work \cite{Zhang_2014} on the non-Markovian effect on the quantum Otto engine indicated that the efficiency of the engine can exceed the Carnot bound, apparently 
violating the second law of thermodynamics. However, subsequent studies \cite{Wiedmann_2020,Shirai_2021,Ishizaki_2023} revealed that 
it was essential to take into account the work required to attach 
and detach the heat baths in order to correctly describe 
the thermodynamics of the engine, highlighting the importance of 
the interaction between the system and the bath.
 
In this paper, we continue the investigation on the effect of the
system-bath interaction in non-Markovian Otto cycles.
We study finite-time quantum Otto cycles whose working medium is a harmonic oscillator.
During isochoric processes, when the oscillator is in contact with a heat bath, it undergoes 
the quantum Brownian motion \cite{grabert_quantum_1988} described by the Caldeira-Leggett model \cite{Caldeira_1983}. The non-Markovian features of this process have been studied 
in Ref.~\cite{einsiedler_non-markovianity_2020}. 
We note that compared to Refs.~\cite{Shirai_2021,Ishizaki_2023} where the working medium
is a single qubit, our system is more complex in the following sense.
For the qubit system with time-dependent level spacing \cite{Shirai_2021,Ishizaki_2023}, 
the time-dependent Hamiltonian commutes with itself at different times, since it
is proportional to a fixed operator. On the other hand, in our model, where
the frequency of the harmonic oscillator changes during adiabatic processes of the cycle, 
the external driving part does not
commute with the Hamiltonian at different times, which is known to be the origin of 
quantum friction \cite{Kosloff_Feldmann_2002,Plastina_2014}.
We study the dynamics of the working medium in contact 
with the heat bath by exactly solving the Heisenberg equations of motion.
Within this approach, we show that we can obtain exact analytic expressions for 
the time evolution of the working medium
and the interaction energy between the system and the bath, which
enables us to explore the parameter space of our model 
and investigate the effect of interaction without much difficulty.
Other approaches to studying non-Markovian Otto cycles use exact master equations \cite{Wiedmann_2020,Wiedmann_2021}, which are known to be computationally costly.

For various values of the parameters describing the hot and cold baths and the coupling
between the system and the bath, we obtain cyclic steady states of the Otto cycle.
We find that the change in the interaction energy during the isochoric process
contributes to both work and heat.
Our calculations show that the change in the interaction energy is always negative,
and its contribution is detrimental to the work output if the cycle operates as an engine. 
In fact, the Otto cycle performs as an engine only when the time in which the system
contacts the heat bath is larger than some value. We also find that the inclusion of the interaction energy is crucial to be consistent with
the thermodynamic laws since there are cases where the efficiency of the engine
exceeds the Carnot efficiency when the interaction energy part is neglected in the calculation
of work and heat. 
An advantage of our approach is that we can take the Markovian approximation mentioned above
on our exact Heisenberg equations and find the corresponding Lindblad equation.
In this way, for given set of parameters, we are able to compare the exact non-Markovian results directly with those of the Markovian approximation. We find that even for small values of
the system-bath coupling constant, our Otto cycle behaves in a 
completely different way than the Markovian counterpart, showing that the interaction 
between the system and the bath plays a crucial role in non-Markovian Otto cycles.
This also confirms that the 
Markovian approximation is valid only in the limit of a vanishingly small coupling constant.
It is important to note that these non-Markovian features
manifest themselves in thermodynamic quantities such as work and heat.
If we look at only the state observables of the working medium, they show a relatively
small difference compared to their Markovian counterparts even at finite coupling.
Only when the effect of the interaction is fully incorporated into work and heat,
non-Markovian effects become significant.

A recent study \cite{Chun_Park_2025} shows that for Markovian Otto engines, one can 
find the maximum power that the engine can achieve for a given efficiency value.
This is known as the power-efficiency trade-off relation, which has originally been discussed 
for {\it classical} thermal engines \cite{Dechant_Sasa_2018}. 
In this paper, we generate the power and efficiency values 
of our non-Markovian Otto engine and investigate whether
the data obey this Markovian power-efficiency bound. We find that 
our data fall far below the Markovian bound. 

The paper is organized as follows. In the next section, we introduce our model 
Hamiltonian for the quantum Otto cycle and present exact expressions 
for the isochoric and adiabatic processes of the cycle. In Sec.~\ref{sec:results},
we present the results of our calculations. 
Finally, we conclude with discussion in the final section.

\section{Quantum Otto Cycle}
\subsection{Model Hamiltonian}

We consider the working medium of a quantum harmonic oscillator of mass $m$ that undergoes 
the quantum Otto cycle consisting of two isochoric and two
adiabatic processes, which will be described in detail in the following. 
The system Hamiltonian of the working medium is given by
\begin{align}
    H_S(t)=\frac{p^{2}}{2 m} +\frac{1}{2} m \omega^{2}(t) x^{2}
    \label{Hs}
\end{align}
with the position and momentum operators $x$ and $p$, respectively.
During isochoric processes, the system with fixed  oscillator frequency $\omega(t)=\omega_h$ ($\omega_c$) 
is in contact with a hot (cold) thermal bath 
at temperature $T_h$ ($T_c$). In this paper, we consider the case where 
the interaction with the thermal bath is
described by the Caldeira-Leggett model \cite{Caldeira_1983}. Each bath is composed of 
a collection of (infinitely many) harmonic oscillators of frequency $\omega_{n,\nu}$ and mass $m_{n,\nu}$, where $\nu=\mathrm{h,c}$ denote the hot and cold baths, respectively.
The Hamiltonian for the thermal baths is given by
\begin{align}
    H^{(\nu)}_B=\sum_{n}\left(\frac{p_{n,\nu}^{2}}{2 m_{n,\nu}} +\frac{1}{2} m_{n,\nu} \omega_{n,\nu}^{2} x_{n,\nu}^{2}\right)
    \label{Hb}
\end{align}
with the usual position and momentum operators $x_{n,\nu}$ and $p_{n,\nu}$, respectively. 
The interaction between the system and the bath $\nu$ is described by the position-position coupling with coupling strength $\kappa_{n,\nu}$ as
\begin{align}
    H^{(\nu)}_I =-x \sum_{n} \kappa_{n,\nu} x_{n,\nu}+x^{2} \sum_{n} 
    \frac{\kappa_{n,\nu}^{2}}{2 m_{n,\nu} \omega_{n,\nu}^{2}},
    \label{Hint}
\end{align}
where the second term, which is known as the counterterm \cite{Caldeira_1983}, ensures 
that the total Hamiltonian $H_{\rm tot}=H_S+H^{(\rm h)}_B+H^{(\rm c)}_B
+H^{(\rm h)}_I+H^{(\rm c)}_I$ 
is positive definite. In the Caldeira-Leggett model, the properties of 
the bath $\nu$ are conveniently described by the spectral function
\begin{align}
J^{(\nu)}(\omega)=\sum_{n} \frac{\kappa_{n,\nu}^{2}}{2 m_{n,\nu} \omega_{n,\nu}} \delta\left(\omega-\omega_{n,\nu}\right) . \label{Jnu}
\end{align}

\begin{figure}[t]
    \centering
    \includegraphics[width=1\linewidth]{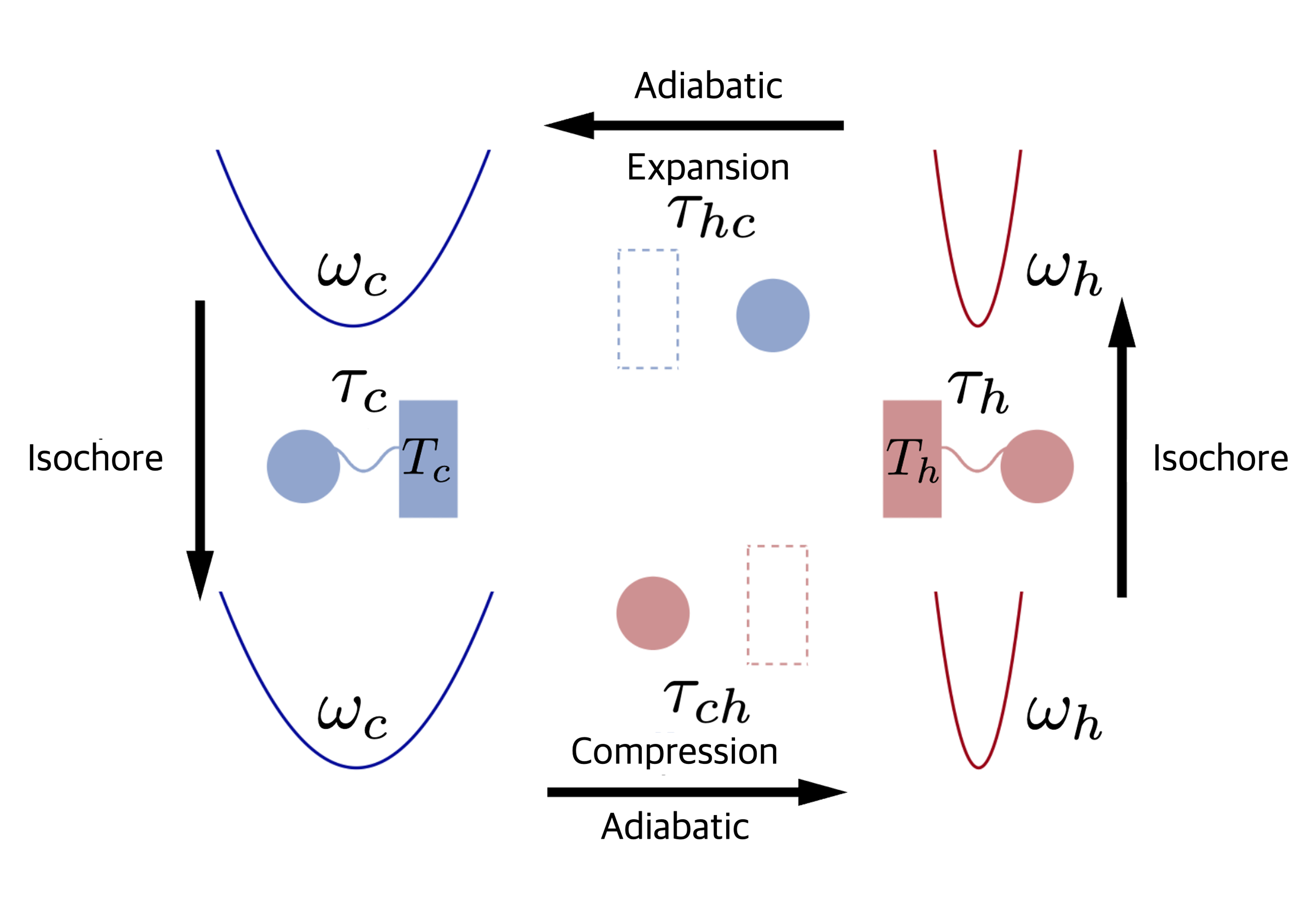}
    \caption{Schematic diagram showing the four stages of the quantum Otto cycle}
    \label{fig:schematic}
\end{figure}

A schematic diagram depicting the quantum Otto cycle we consider is shown in Fig.~\ref{fig:schematic}. At the beginning of the hot isochoric process, the system with the oscillator frequency
$\omega_h$ is attached to the hot bath at temperature $T_h$
described by the density operator ($\nu=\mathrm{h}$) 
\begin{align}
\rho^{(\nu)}_B=\frac 1 {Z^{(\nu)}_B}\exp(-H^{(\nu)}_B/T_\nu)
\label{rhob}
\end{align}
with $Z^{(\nu)}_B
=\mathrm{Tr}_B \exp(-H^{(\nu)}_B/T_\nu) $ ($k_B=1$). The total system plus bath 
evolves unitarily for the duration of $\tau_h$. At the end of the hot isochoric process,
we determine 
the system properties by taking the partial trace over the bath variables
as we will describe in the next subsection, and detach the bath from the system.  
The system then undergoes an adiabatic expansion where $\omega(t)$ changes from $\omega_h$
to $\omega_c(<\omega_h)$ following a predetermined protocol for the duration $\tau_{hc}$.
Subsequently, the system is attached to the cold bath at temperature $T_c(<T_h)$ and undergoes the isochoric process similar to the one explained above for the duration $\tau_c$. After the system is detached from the cold bath, the final adiabatic compression of
the oscillator frequency of the system from $\omega_c$ to $\omega_h$ occurs during the time period $\tau_{ch}$ and completes the cycle.

In the following two subsections, we investigate how the system evolves in time during each process
and show how the cyclic steady state can be found. 

\subsection{Time Evolution in the Isochoric Process}
\label{sec:isochore}

For the isochoric process, we 
need to look at the time evolution of an open quantum system. In many previous works 
\cite{Rezek_Kosloff_2006, Kosloff_Rezek_2017, lee2020, Chun_Park_2025}
on the quantum Otto cycle, a Markovian bath is usually assumed, where the dynamics
is governed by the Lindblad equation. In this paper, we consider the exact Heisenberg-Langevin equations for the system observables, which does not rely on the weak coupling 
or the Markovian assumption. We can set up Heisenberg equations for the Heisenberg operators
$x(t)$, $p(t)$, $x_{n,\nu}(t)$ and $p_{n,\nu}(t)$ from Eqs.~(\ref{Hs}), (\ref{Hb}) and (\ref{Hint}). After solving the equations for the bath operators and inserting back into 
the system equations, we get $\dot{x}(t)=p(t)/m$ and for $\nu=\mathrm{h,c}$
\begin{align}
\ddot x(t)+\omega_\nu^{2} x(t)+\frac{d}{d t} \int_{0}^{t} d t^{\prime} \gamma^{(\nu)}\left(t-t^{\prime}\right) x\left(t^{\prime}\right)=\frac{1}{m} B^{(\nu)}(t), \label{EOM}
\end{align}
where
 \begin{align}
\gamma^{(\nu)}(t)=\frac{2}{m} \int_{0}^{\infty} d \omega \frac{J^{(\nu)}(\omega)}{\omega} \cos (\omega t) \label{gamma}
\end{align}
is the dissipation kernel 
and 
\begin{align}
B^{(\nu)}(t)\equiv \sum_n \kappa_{n,\nu} &\Big(x_{n,\nu}(0)\cos(\omega_{n,\nu} t)  \nonumber \\
&+\frac{p_{n,\nu}(0)}{m_{n,\nu}\omega_{n,\nu}}\sin(\omega_{n,\nu} t) \Big)
\label{Bath}
\end{align}
is the bath operator determined by the initial state of the bath. 
This equation can be expressed in terms of homogeneous solutions,
$G^{(\nu)}_1(t)$ and $G^{(\nu)}_2(t)$ to Eq.~(\ref{EOM}) as
 \begin{align}
x(t)=&G^{(\nu)}_{1}(t) x(0)+G^{(\nu)}_{2}(t) \frac{p(0)}{m}
\nonumber \\
&+\frac{1}{m} \int_{0}^{t} d t^{\prime} G^{(\nu)}_{2}\left(t-t^{\prime}\right) B^{(\nu)}(t^{\prime}) \label{xsol},
\end{align}
where the initial conditions are given by
$G^{(\nu)}_1(0)=1$, $\dot{G}^{(\nu)}_1(0)=0$, $G^{(\nu)}_2(0)=0$
and $\dot{G}^{(\nu)}_2(0)=1$.

At the beginning of the isochoric process, the system is attached to the bath $\nu$.
The initial state of the total system is $\rho_S(0)\otimes \rho^{(\nu)}_B$ for some
initial system state $\rho_S(0)$ and the equilibrium state as in Eq.~(\ref{rhob}) for the bath. 
At later times $t>0$, the total system evolves from this initial product state.
In the Heisenberg picture, the time evolution is encoded in the time dependence of 
the operator as in Eq.~(\ref{xsol}) and 
the averages are calculated with respect to the initial state. 

To study the properties of the Otto cycle, it is sufficient to 
monitor the averages of the following quantities of the system variables \cite{Rezek_Kosloff_2006,Kosloff_Rezek_2017}:
\begin{align}
   &\langle H(t)\rangle\equiv \frac 1 {2m} \langle p^2(t)\rangle +
    \frac 1 2 m \omega^2_\nu \langle x^2(t)\rangle , \label{Ht:def}\\
&\langle L(t)\rangle\equiv \frac 1 {2m} \langle p^2(t)\rangle -
    \frac 1 2 m \omega^2_\nu \langle x^2(t)\rangle  , \label{Lt:def}
    \end{align}
    and
    \begin{align}
    \langle C(t)\rangle\equiv \frac {\omega_\nu} 2 \langle  x(t) p(t)+p(t) x(t)\rangle 
    \label{Ct:def}
\end{align}
for each isochoric process $\nu=\mathrm{h,c}$,
where the average $\langle\cdots\rangle$ is evaluated with respect to the
initial product state mentioned above. 
Equation (10) of course monitors the average energy of the system.
In order to specify the state of the system at an arbitrary point in the Otto cycle,
it is convenient to introduce a column vector \cite{Rezek_Kosloff_2006,Kosloff_Rezek_2017}
\begin{align}
    \phi(t) =(\langle H(t)\rangle,\langle L(t)\rangle,\langle C(t)\rangle,1)^{\mathrm{T}}.
    \label{phit}
\end{align}
Using Eq.~(\ref{xsol}) and $p(t)=m\dot{x}(t)$, we can express 
$\phi(t)$ 
in terms of $\phi(0)$
and the averages of $B^{(\nu)}(t)$ with respect to $\rho^{(\nu)}_B$. For the latter average,
we have $\langle B^{(\nu)}(t)\rangle=0$ and
\begin{align}
    D^{(\nu)}_1(t-s)&\equiv  \langle \{ B^{(\nu)}(t),B^{(\nu)}(s)\}\rangle \label{D_1} \\
    =&2\int_0^\infty d\omega\; J^{(\nu)}(\omega)\coth\left(\frac{\omega}{2 T_\nu}\right)\cos(\omega(t-s)) \nonumber 
\end{align}

After some algebra, we have the following.
\begin{widetext}
\begin{align}
    \langle H(t)\rangle =&\frac 1 2 \left[ \frac{1}{\omega^2_\nu}
    \dot G_1^{(\nu)2}(t) + 2G_{1}^{(\nu)2}(t) + \omega^2_\nu G_{2}^{(\nu)2}(t) \right] \langle H(0)\rangle 
    +   \frac 1 2 \left[- \frac{1}{\omega^2_\nu}\dot G_1^{(\nu)2}(t)  + \omega^2_\nu G_{2}^{(\nu)2}(t) \right] \langle L(0)\rangle  \notag\\
   + & \frac{1}{\omega_\nu}\left[\dot G_1^{(\nu)}(t)\dot G_2^{(\nu)}(t)+\omega^2_\nu G_{1}^{(\nu)}(t)G_{2}^{(\nu)}(t) \right] \langle C(0)\rangle + \langle I^{(\nu)}_H(t)\rangle , 
   \label{Ht}
\end{align}
\begin{align}
    \langle L(t)\rangle =&\frac 1 2 \left[ \frac{1}{\omega^2_\nu}\dot G_1^{(\nu)2}(t)  - \omega^2_\nu G_{2}^{(\nu)}(t) \right] \langle H(0)\rangle 
+    \frac 1 2 \left[- \frac{1}{\omega^2_\nu}\dot G_1^{(\nu)2}(t) 
+2 G_{1}^{(\nu)2}(t)  - \omega^2_\nu G_{2}^{(\nu)2}(t) \right] \langle L(0)\rangle \notag \\
   + & \frac{1}{\omega_\nu} \left[\dot G_1^{(\nu)}(t)\dot G_2^{(\nu)}(t)-\omega^2_\nu G_{1}^{(\nu)}(t)G_{2}^{(\nu)}(t) \right] \langle C(0)\rangle +\langle I^{(\nu)}_L(t)\rangle,
   \label{Lt}
\end{align}
and
\begin{align}
    \langle C(t)\rangle
    &= \left[ \frac{1}{\omega_\nu}\dot G_1^{(\nu)}(t)G_1^{(\nu)}(t)
    +\omega_\nu\dot G_2^{(\nu)}(t)G_2^{(\nu)}(t)\right]\langle H(0)\rangle
   + \left[ -\frac{1}{\omega_\nu}\dot G_1^{(\nu)}(t)G_1^{(\nu)}(t)
    +\dot G_2^{(\nu)}(t)G_2^{(\nu)}(t)\right]\langle L(0)\rangle \notag \\
    &+(\dot G_1^{(\nu)}(t)G_2^{(\nu)}(t)+\dot G_2^{(\nu)}(t)G_1^{(\nu)}(t))\langle C(0)\rangle+\langle I_C^{(\nu)}(t)\rangle,
    \label{Ct}
\end{align}
\end{widetext}
where $\langle I^{(\nu)}_H(t)\rangle ,\langle I^{(\nu)}_L(t)\rangle$ and
$\langle I^{(\nu)}_C(t)\rangle$ are obtained from the averages of the bath variables.
They are given by
\begin{align}
&    \langle I^{(\nu)}_H(t)\rangle =\frac{1}{2m} \langle I^{(\nu)}_{pp}(t)\rangle+\frac m 2 \omega^2_\nu \langle I^{(\nu)}_{xx}(t)\rangle ,\\
&\langle I^{(\nu)}_L(t)\rangle =    \frac{1}{2m} \langle I^{(\nu)}_{pp}(t)\rangle-\frac m 2 \omega^2_\nu \langle I^{(\nu)}_{xx}(t)\rangle
\end{align}
with
\begin{align}
    \langle I^{(\nu)}_{xx}(t)\rangle=\frac{1}{2m^2}&\int_0^t ds \int_0^t ds^\prime 
    G^{(\nu)}_2(t-s) \notag \\
    &\times G^{(\nu)}_2(t-s^\prime) D^{(\nu)}_1(s-s^\prime ), \label{Ixx}
\end{align}
\begin{align}
     \langle I^{(\nu)}_{pp}(t)\rangle=\frac{1}{2}&\int_0^t ds \int_0^t ds^\prime 
    \dot{G}^{(\nu)}_2(t-s) \notag \\
    &\times \dot{G}^{(\nu)}_2(t-s^\prime) D^{(\nu)}_1(s-s^\prime ), \label{Ipp}
\end{align}
and
\begin{align}
    \langle I^{(\nu)}_C(t)\rangle  =\frac{\omega_\nu}{2m} &
   \int_0^t ds \int_0^t ds^\prime 
    \dot{G}^{(\nu)}_2(t-s) \notag \\
    &\times G^{(\nu)}_2(t-s^\prime) D^{(\nu)}_1(s-s^\prime ) . \label{IC}
\end{align}
From Eqs.~(\ref{Ht}), (\ref{Lt}) and (\ref{Ct}), it is straightforward to construct
the $4\times 4$ matrix $\mathcal{P}^{(\nu)}(t)$ such that the state of the system $\phi(t)$ in the isochoric process
at time $t$ after the system is in contact with the thermal bath $\nu$ at time $t=0$ is
given by 
\begin{align}
    \phi(t)=\mathcal{P}^{(\nu)}_{\rm iso}(t) \phi(0).
    \label{prop:iso}
\end{align}

The isochoric propagator $\mathcal{P}^{(\nu)}_{\rm iso}(t)$ can be evaluated once the 
spectral density in Eq.~(\ref{Jnu}) for the bath is given. 
In this paper, we take the simple Ohmic form with
the Lorentz-Drude cutoff as
\begin{align}
J^{(\nu)}(\omega)=\frac{2 m \gamma}{\pi} \omega \frac{\Omega^{2}}{\Omega^{2}+\omega^{2}} ,\label{J}
\end{align}
where $\gamma$ gives the coupling strength between the system and the bath,
and $\Omega$ is the cutoff frequency. We note that for simplicity 
we take the same values of $\gamma$ and $\Omega$ for both hot and cold baths.
The detailed expression for $\mathcal{P}^{(\nu)}_{\rm iso}(t)$ for the 
Ohmic spectral density with the Lorentz-Drude cutoff is given in Appendix \ref{app:1}.


\subsection{The Adiabatic Process and Cyclic Steady States}
In the adiabatic processes of the Otto cycle, the system is detached from the bath and the 
frequency of the oscillator $\omega(t)$ changes from $\omega_i$ to
$\omega_f$. 
As in Eqs.~(\ref{Ht:def}), (\ref{Lt:def} and (\ref{Ct:def}) for the isochoric process,
we monitor the state of the system through
the time dependence of the system Hamiltonian $H(t)$ 
(we drop the subscript $S$ in this process), the system Lagrangian $L(t)$
and the anticommutator
of the position and momentum $C(t)$ defined as
\begin{align}
&    H(t)=\frac{p^2(t)}{2m}+\frac 1 2 m\omega^2(t)x^2(t), \\
&    L(t)=\frac{p^2(t)}{2m}-\frac 1 2 m\omega^2(t)x^2(t), \\
&    C(t)=\frac{\omega(t)}{2}\left[ x(t) p(t)+p(t) x(t)\right].
\end{align}
The time dependence of any system operator $O(t)$ is governed by
the system Hamiltonian $H(t)$ via the Heisenberg equation of motion. 
They are given by
\begin{align}
    \frac{d}{dt} \begin{pmatrix}
  H(t)\\
  L(t)\\
 C(t) 
\end{pmatrix}=\omega(t)\begin{pmatrix}
\mu & -\mu & 0  \\
-\mu & \mu & -2  \\
0 & 2 & \mu \\
\end{pmatrix} \begin{pmatrix}
  H(t)\\
  L(t)\\
 C(t) 
\end{pmatrix} ,
\label{matrixeq}
\end{align}
where
\begin{align}
    \mu(t)\equiv \frac{\dot\omega(t)}{\omega^2(t)}
\end{align}
is, in general, a function of time. In many previous studies of quantum Otto 
engines \cite{Kosloff_Rezek_2017, lee2020}, a simple case where $\mu(t)=\mu$ is 
a constant has been studied. In this paper, we will also use this simple
protocol mainly because it allows an analytical expression for the adiabatic propagator
given below. For this protocol, if we start from $\omega(0)=\omega_i$ 
and end at $\omega(\tau)=\omega_f$, the time dependence of the oscillator frequency is
given by
\begin{align}
    \omega(t)=\frac{\omega_i\omega_f}{\omega_f-(\omega_f-\omega_i)t/\tau}.
    \label{protocol}
\end{align}
Given the set of parameters $(\omega_i,\omega_f,\tau)$, $\mu$ is determined by
$\mu\tau=\omega_i^{-1}-\omega_f^{-1}$.
In our Otto cycle, we have two adiabatic processes
characterized by the set of parameters $(\omega_i,\omega_f,\tau)$ given by
$(\omega_c,\omega_h,\tau_{ch})$ and $(\omega_h, \omega_c,\tau_{hc})$.
The former is referred to as an adiabatic compression, and the latter is an adiabatic expansion.

Now, from the equation of motion for the system operators, Eq.~(\ref{matrixeq}),
it is straightforward
to derive the expressions for the propagator
$\mathcal{P}_{\mathrm{ad}}(\tau)$ in the adiabatic process, 
which describes the time evolution of the 
state $\phi(t)$ as given in Eq.~(\ref{phit}) 
from the beginning of the process at time $t=0$ to 
the end at $\tau>0$: $\phi(\tau)=\mathcal{P}_{\rm ad}(\tau)\phi(0)$. 
The detailed expressions for the propagators 
$\mathcal{P}^{\rm (c\to h)}_{\rm ad}(\tau_{ch})$
and $\mathcal{P}^{\rm (h\to c)}_{\rm ad}(\tau_{hc})$ for the two adiabatic processes 
are given in Appendix \ref{app:2}. 

The complete Otto cycle consists of combining two isochoric and two adiabatic processes as  
described in Fig.~\ref{fig:schematic}. The propagator for a cycle starting from
the adiabatic compression ($\omega_c\to\omega_h$) is given by
\begin{align}
    \mathcal{P}_{\rm cyc}\equiv \mathcal{P}_{\rm iso}^{\rm (c)}(\tau_c)
    \mathcal{P}_{\rm ad}^{\rm (h\to c)}(\tau_{hc})
    \mathcal{P}_{\rm iso}^{\rm (h)}(\tau_h)\mathcal{P}_{\rm ad}^{\rm(c\to h)}(\tau_{ch}),
    \label{P_cyc}
\end{align}
where the system is in contact
with the hot and cold baths for a duration $\tau_h$ and $\tau_c$, respectively.
In the following, we focus on the cyclic steady state $\phi_{\rm ss}$ that satisfies
\begin{align}
    \phi_{\rm ss}=\mathcal{P}_{\rm cyc} \phi_{\rm ss}.
    \label{steady}
\end{align}

\subsection{Work, Heat and Interaction Energy}

Once a cyclic steady state is found, we can investigate the energy change in
each process as follows. The system energy changes $\langle \Delta H_S\rangle_{ch}$ and
$\langle \Delta H_S\rangle_{hc}$
during the adiabatic processes with the change in frequency, $\omega_c\to\omega_h$ 
and $\omega_h\to\omega_c$, respectively, can be calculated as
\begin{align}
    \langle \Delta H_S\rangle_{ch}=&\Pi_1(\mathcal{P}^{\rm (c\to h)}_{\rm ad}-
    \mathcal{I})\phi_{ss},\\
     \langle \Delta H_S\rangle_{hc}=&
    \Pi_1 (\mathcal{P}_{\rm ad}^{\rm (h\to c)}-\mathcal{I})\mathcal{P}_{\rm iso}^{\rm (h)}\mathcal{P}_{\rm ad}^{\rm(c\to h)}\phi_{ss},
    \end{align}
    where $\mathcal{I}$ is the $4\times 4$ identity matrix and
    \begin{align}
    \Pi_1 \equiv (1,0,0,0)
\end{align}
is a row vector that selects the first element of the state vector (i.e.\ the energy).  
Similarly, we can calculate the system energy changes $\langle \Delta H_S\rangle_{h}$ 
and $\langle \Delta H_S\rangle_{c}$ during the isochoric processes in contact with 
the hot and cold bath, respectively, as
    \begin{align}
    \langle \Delta H_S\rangle_{h}=&\Pi_1(
    \mathcal{P}_{\rm iso}^{\rm (h)}
    -\mathcal{I})\mathcal{P}^{\rm (c\to h)}_{\rm ad}
    \phi_{ss}, \\
    \langle \Delta H_S\rangle_{c}=&\Pi_1(\mathcal{P}_{\rm iso}^{\rm (c)}-\mathcal{I})
    \mathcal{P}_{\rm ad}^{\rm (h\to c)}
    \mathcal{P}_{\rm iso}^{\rm (h)}\mathcal{P}_{\rm ad}^{\rm(c\to h)}
    \phi_{ss}.
\end{align}

If we can somehow ignore the interaction energy between the system and the bath, 
we may associate the energy change in the adiabatic process
with the work $\widetilde{W}$ 
done \textit{by} the system as
\begin{align}
    \widetilde{W}= -  \langle \Delta H_S\rangle_{ch} - \langle \Delta H_S\rangle_{hc}.
    \label{tildew}
\end{align}
Similarly, the energy changes during the isochoric process 
can be regarded as the heat  $\widetilde{Q}_h$ and $\widetilde{Q}_c$ absorbed 
by the system from the hot and cold baths, respectively:
\begin{align}
    \widetilde{Q}_h=\langle \Delta H_S\rangle_{h},~~~\widetilde{Q}_c
    =\langle \Delta H_S\rangle_{c}.
    \label{tildeq}
\end{align}
Since these are for a cyclic steady state, we have 
\begin{align}
    \widetilde{W}=\widetilde{Q}_h+\widetilde{Q}_c
    \label{1st:markov}
\end{align}

In the present non-Markovian case, however, the interaction energy cannot be neglected. 
The effect of the interaction energy on non-Markovian quantum engines
has been studied previously \cite{Wiedmann_2020,Wiedmann_2021,Shirai_2021,Ishizaki_2023}.
For the work done by the system, in addition to the energy change in the adiabatic process, the cost of energy involved in attaching and detaching the heat baths must be included.
To be more concrete, the work done by the system is defined as \cite{Wiedmann_2020}
\begin{align}
    W=-\int_0^{\tau_{\rm cyc}} dt \;  \left\langle \frac{\partial H_{\rm tot}(t)}{\partial t} \right\rangle ,
    \label{work}
\end{align}
where $\tau_{\rm cyc}=\tau_{ch}+\tau_h+\tau_{hc}+\tau_c$.
The explicit time-dependence of the total Hamiltonian comes from two sources. 
The first one is 
the adiabatic processes where the frequency of the oscillator changes
considered above. The contribution 
of this part to Eq.~(\ref{work}) is exactly $\widetilde{W}$ in Eq.~(\ref{tildew}).
The second part is the interaction Hamiltonian $H_I^{(\nu)}$ in Eq.~(\ref{Hint}), which 
is nonzero only for the period of the isochoric processes and vanishes otherwise. Therefore,
we may regard $H_I^{(\nu)}$ as having a factor of 
step functions in time in front of it. For example, if 
the isochoric process with the bath $\nu$ starts from $t=t_\nu$ and ends at $t=t_\nu+\tau_\nu$,
then $H^{(\nu)}_I$ contains a factor of $\theta(t-t_\nu) - \theta(t-t_\nu-\tau_\nu)$, 
where $\theta(x)=0$, for $x<0$ and $\theta(x)=1$ for $x>0$ is the step function.
The time derivative of this factor gives us $\delta(t-t_\nu)-\delta(t-t_\nu-\tau_\nu)$.
Therefore, the contribution of this part to the work in Eq.~(\ref{work}) is the change in the
interaction energy during the isochoric processes and can be interpreted as the work involved in
attaching and detaching the heat baths. Combining these two contributions, we can write
\begin{align}
    W=\widetilde{W}+\langle \Delta H^{\rm (h)}_I\rangle +\langle \Delta H^{\rm (c)}_I\rangle ,
    \label{work_int}
\end{align}
where
\begin{align}
    \langle \Delta H^{(\nu)}_I\rangle=\langle H^{(h)}_I (t_\nu+\tau_\nu)\rangle
    -\langle  H^{(h)}_I (t_\nu)\rangle
    \label{DeltaHI}
\end{align}
is the change in the interaction energy during 
the isochoric process with the bath $\nu=\mathrm{h,c}$.

The heat absorbed by the system must be modified due to the presence of the interaction Hamiltonian. In order to have the first law 
\begin{align}
    W=Q_h+Q_c ,
\end{align}
as a generalization of Eq.~(\ref{1st:markov}), it is natural to define
the heat $Q_h$ and $Q_c$ absorbed from the hot and cold bath, respectively, as
\begin{align}
&    Q_h=\widetilde{Q}_h+\langle \Delta H^{\rm (h)}_I\rangle , \label{qh} \\
&     Q_c=\widetilde{Q}_c+\langle \Delta H^{\rm (c)}_I\rangle . \label{qc}
\end{align}
The heat defined in this way is, in fact, the negative of the change in the bath energy.
Indeed, from the Heisenberg equation of motion, we have
\begin{align}
    -\frac{d H^{(\nu)}_{\rm B}}{dt} =& -i[H_{\rm tot},H^{(\nu)}_{\rm B}] \\
    =& i [H_{\rm tot},H_{\rm S}]+i[H_{\rm tot},H^{(\nu)}_{\rm I}].
\end{align}
When the average of this quantity is integrated over the isochoric process, the first term 
gives the system energy change $\widetilde{Q}_{\nu}$ and the second term
$\langle \Delta H^{(\nu)}_I\rangle$.

The average interaction energy and its change can be calculated as follows. 
Since the system and bath operators commute each other, we can write from Eq.~(\ref{Hint})
    \begin{align}
        \langle H^{(\nu)}_I(t)\rangle =& -\frac 1 2 \left\langle 
        \left\{ 
        x(t),\sum_n\kappa_{n,\nu} x_{n,\nu} (t)
        \right\}
        \right\rangle \notag\\
        &+\frac{\mu^{(\nu)}}{2} \left\langle x^2(t) \right\rangle,
        \label{HInu}
        \end{align}
where 
\begin{align}
    \mu^{(\nu)}=\sum_n \frac{\kappa^2_{n,\nu}}{m_{n,\nu} \omega^2_{n,\nu}} =m\gamma^{(\nu)}(0).
    \label{mu}
\end{align}
Using the solution to
the Heisenberg equation of motion for the bath variable, we can write
\begin{align}
    \sum_n \kappa_{n,\nu}x_{n,\nu}(t)&=B^{(\nu)}(t) \notag\\
    &-m\int_0^t ds\; \eta^{(\nu)}(t-s)x(s) ,
    \label{xnsol}
\end{align}
where
\begin{align}
    \eta^{(\nu)}(t)=\frac{d}{dt}\gamma^{(\nu)}(t)=-\frac{2}{m}\int_0^\infty d\omega\;
    J^{(\nu)}(\omega)\sin(\omega t).
    \label{etat}
\end{align}
Combining Eqs.~(\ref{HInu}) and (\ref{xnsol}), we have
        \begin{align}
        &\langle H^{(\nu)}_I(t)\rangle 
        =-\frac 1 2 \left\langle 
        \left\{
        x(t), B^{(\nu)}(t) \right\}\right\rangle  \label{HInu_sol}\\
        & ~~+ \frac{m}{2}\int_0^t ds\; \eta^{(\nu)}(t-s) \left\langle \left\{ x(t), x(s)
        \right\} 
        \right\rangle +\frac{\mu^{(\nu)}}{2} \left\langle x^2(t) \right\rangle .\notag
    \end{align}
We can then evaluate $\langle\Delta H^{(\nu)}_I \rangle$ using the solution
Eq.~(\ref{xsol}). The calculation leading to this is rather involved and 
is given in detail in Appendix \ref{app:3}.

\subsection{The Markovian Limit}
One of the purposes of this paper is to compare
our results with those of the Otto cycle operated in the Markovian limit, 
which has been studied extensively before \cite{Feldmann_Kosloff_2004,Rezek_Kosloff_2006,Agarwal_Chaturvedi_2013,Zheng_Poletti_2014,Kosloff_Rezek_2017,Insinga_2018,Kloc_Cejnar_Schaller_2019,Abah_Paternostro_2019,Park_Lee_Chun_Noh_2019,Chen_Sun_Dong_2019,Dann_Kosloff_Salamon_2020,lee2020}.
In the Markovian Otto cycle, 
the isochoric processes are completely governed by the 
Lindblad equation. 
The reduced density 
operator $\rho_S(t)$ in the Schr\"odinger picture
for the system in contact with the bath $\nu=\mathrm{h,c}$ is given by
\begin{align}
    \frac{d}{dt}\rho_S(t)&=-i[H^{(\nu)}_S,\rho_S(t)] \notag \\
    &+\gamma^{(\nu)}_L(\bar{n}_\nu+1)\left(a_\nu\rho_S a^\dagger_\nu-\frac 1 2\{ a^\dagger_\nu
    a_\nu,\rho_S\}\right) \notag\\
    &+\gamma^{(\nu)}_L\bar{n}_\nu\left(a^\dagger_\nu\rho_S a_\nu-\frac 1 2\{ a_\nu
    a^\dagger_\nu,\rho_S\}\right) ,
    \label{lindblad}
    \end{align}
where $H^{(\nu)}_S$ is the system Hamiltonian, Eq.~(\ref{Hs}) with $\omega(t)=\omega_\nu$, $\bar{n}_\nu=(e^{\omega_\nu/T_\nu}-1)^{-1}$,
and $a_\nu=\sqrt{m\omega_\nu/2}(x+ip/(m\omega_\nu))$.
This can be obtained from 
the master equation corresponding to the
exact Heisenberg equations of motions we are using 
by taking a series of well-known approximations, namely the Born-Markov
followed by the secular approximations \cite{OQS_Breuer}. 
The parameter $\gamma^{(\nu)}_L$ in Eq.~(\ref{lindblad}) is given by
\cite{OQS_Breuer, Picatoste_2024}
\begin{align}
\gamma^{(\nu)}_L=    \frac{2\pi J^{(\nu)}(\omega_\nu)}{2m\omega_\nu}
=2\gamma\frac{\Omega^2}{\omega^2_\nu+\Omega^2}.
\label{eff_gamma}
\end{align}
Therefore, in the Markovian limit,
our model for isochores corresponds to 
the Lindblad equation with different dissipation parameters
$\gamma^{(\nu)}_L$
for hot and cold baths. The isochoric propagators defined as in 
Eq.~(\ref{prop:iso}) in this case have been studied in many previous works (see, for example
Refs.~\cite{Kosloff_Rezek_2017,lee2020}).

\section{Results}
\label{sec:results}

In this section, we present in detail the results of calculations
of the physical quantities arising in the Otto cycle discussed in the previous section. 
The main quantities involved in the isochoric and adiabatic processes are
the propagators given in Eqs.~(\ref{Piso}) and (\ref{padia}). 
We evaluate these propagators for a given set of parameters,
$T_h$, $T_c$, $\omega_c$ and $\omega_h$ that 
describe the Otto cycle and the heat baths.
As explained in the previous section, we need to 
specify the times during which the isochoric and adiabatic processes
are performed. For simplicity, we only consider the case 
where $\tau_{hc}=\tau_{ch}$ and $\tau_h=\tau_c$. 
Additional parameters we need are the coupling strength $\gamma$
between the system and the bath and the cutoff frequency $\Omega$
of the baths as given by Eq.~(\ref{J}). We use the unit where
$k_B=\hbar=1$ and $m=1$.

From the propagators, we look for cyclic
steady states that the Otto cycle can reach. 
These steady states are characterized by Eq.~(\ref{steady}).
We therefore look for cases where the 
largest eigenvalue of $\mathcal{P}_{\rm cyc}$
is 1. The other eigenvalues will determine the rate of approach to steady state \cite{Kosloff_Rezek_2017}. 
Due to the particular structure of $\mathcal{P}_{\rm cyc}$
whose fourth row is given by $(0,0,0,1)$, there is always an eigenvalue equal to one. 
If the eigenvector corresponding to the eigenvalue one has a non-vanishing
fourth element, we can normalize it to one to get the cyclic steady state  $\phi_{\rm ss}$.
When the fourth element of the eigenvector corresponding to the unit eigenvalue is zero, 
this does not correspond to a physical state (see Eq.~(\ref{phit})),
which signals the absence of the cyclic steady state.
The same criterion for the existence of steady states has been used for 
the Markovian Otto cycles \cite{Insinga_2018}. Since the Markovian propagator matrix
has the same structure (with $(0,0,0,1)$ as the fourth row)
as our $\mathcal{P}_{\rm cyc}$, we can apply the same method here.

\begin{figure}[t]
   \centering
    \includegraphics[width=1\linewidth]{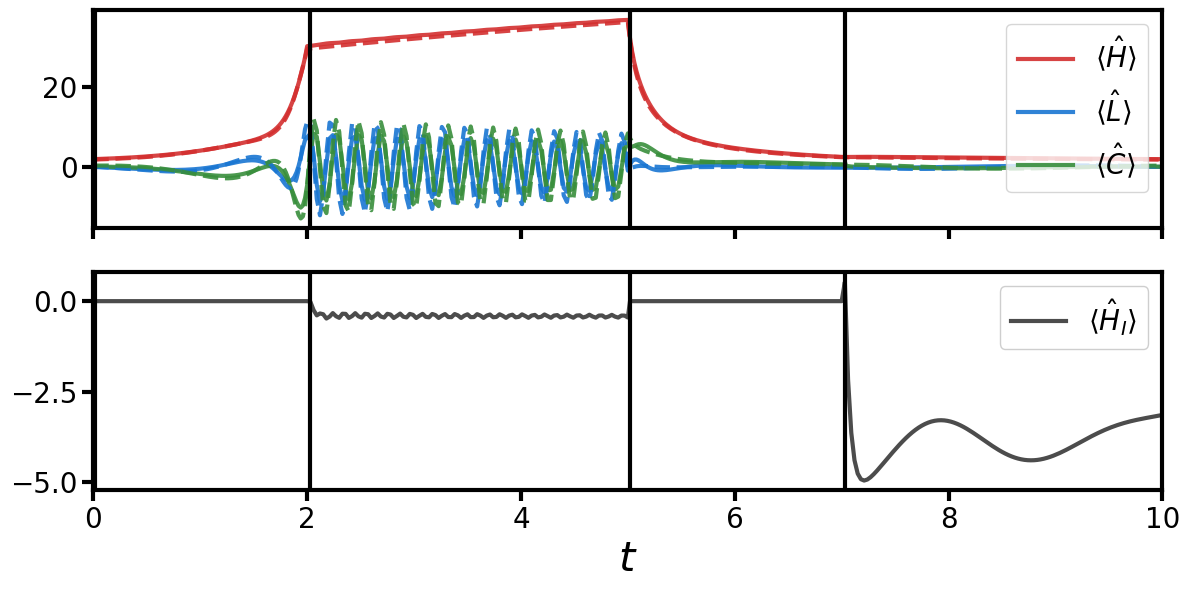}
    \caption{The upper panel shows the time evolution of the state variables 
    $\langle H\rangle$ (red), $\langle L\rangle$ (blue) and $\langle C\rangle$ (green) 
    of the periodic steady state for the Otto
    cycle which consists of the adiabatic processes with $\tau_h=\tau_c=3$ and the isochoric ones with $\tau_{hc}=\tau_{ch}=2$ (see Eq.~(\ref{P_cyc}). The parameters used are
    $\omega_c=1$, $\omega_h=15$, $T_c=1$, $T_h=50$, $\Omega=20$ and $\gamma=0.1$.
    The solid lines are results from the exact solutions of the Heisenberg-Langevin equations and the dashed lines are from the Lindblad equation. 
    The lower panel shows the time evolution of the average interaction energy
    $\langle H_I(t)\rangle$ during the same Otto cycle.}
    \label{fig:evolution1}
\end{figure}

\begin{figure}[b]
    \centering
    \includegraphics[width=1\linewidth]{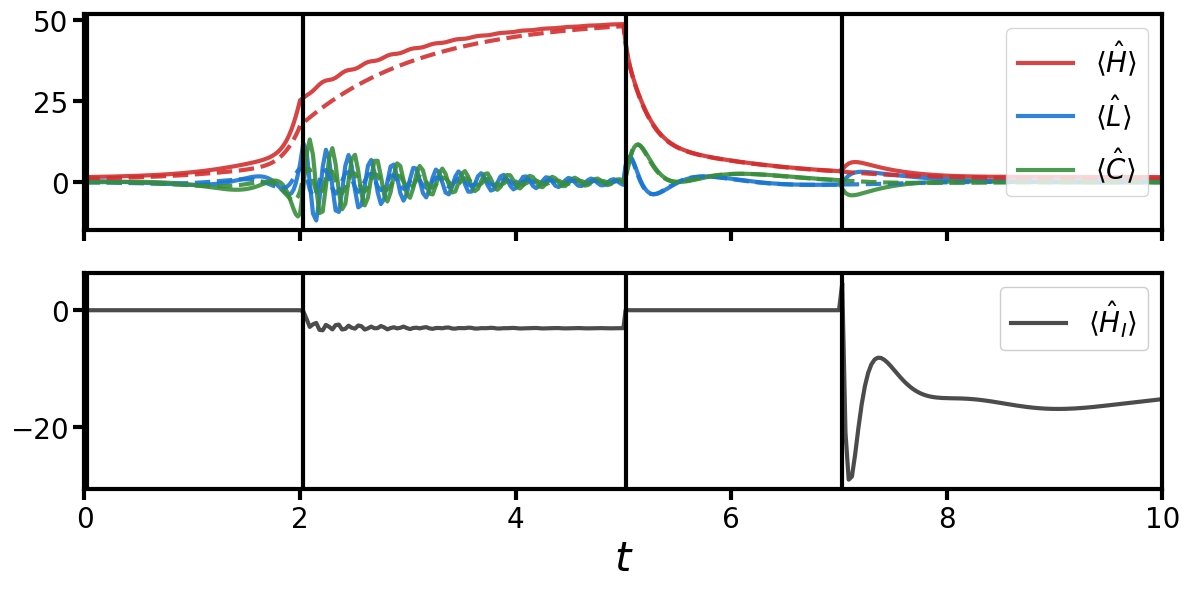}
    \caption{Same as Fig.~\ref{fig:evolution1} except for the coupling strength
    $\gamma=0.7$ between the system and the bath.}
    \label{fig:evolution2}
\end{figure}

The upper panels of Figs.~\ref{fig:evolution1} and \ref{fig:evolution2} show
the time evolution of the system variables $\phi(t)$ in Eq.~(\ref{phit})
for the typical cyclic steady states we found. 
We show the results obtained by solving the exact Heisenberg equations 
in Eq.~(\ref{Piso}) (solid lines) along with those of the Markovian approximation
(dashed lines) obtained from Eqs.~(\ref{lindblad}) and (\ref{eff_gamma}).
As expected, when the coupling strength $\gamma$ between the system 
and the bath is weak (see Fig.~\ref{fig:evolution1}), the two steady states
are quite similar. On the other hand, as the coupling becomes strong, 
our results from the exact Heisenberg equations start to deviate from 
the Markovian approximation results,
as can be seen from Fig.~\ref{fig:evolution2}. 
We note, however, that the differences of these state variables of
the working medium from their Markovian counterparts are not that significant  
even for finite coupling strength. As we show in the following, 
more significant non-Markovian 
effects appear when we consider thermodynamic fluxes such as work and heat,
for which the system-bath interaction is taken into account. 

We now consider the effect of the interaction energy $H_I$ between the system and the bath during the 
isochoric process. In the Markovian approximation, by its construction, the interaction energy 
is completely ignored. On the other hand, in our exact Heisenberg equation approach,
the time evolution of the interaction energy during the isochoric process of cyclic steady states
can be calculated exactly as given in Appendix \ref{app:3}. Examples of this calculation are shown 
in the lower panels of Figs.~\ref{fig:evolution1} and \ref{fig:evolution2}.
We note that in most cases we have studied
the change in the interaction energy $\langle \Delta H^{(\nu)}_I\rangle$ 
between the beginning and end of the isochoric process is negative for both hot and cold baths.

\begin{figure}[b]
    \centering
   \includegraphics[width=0.8\linewidth]{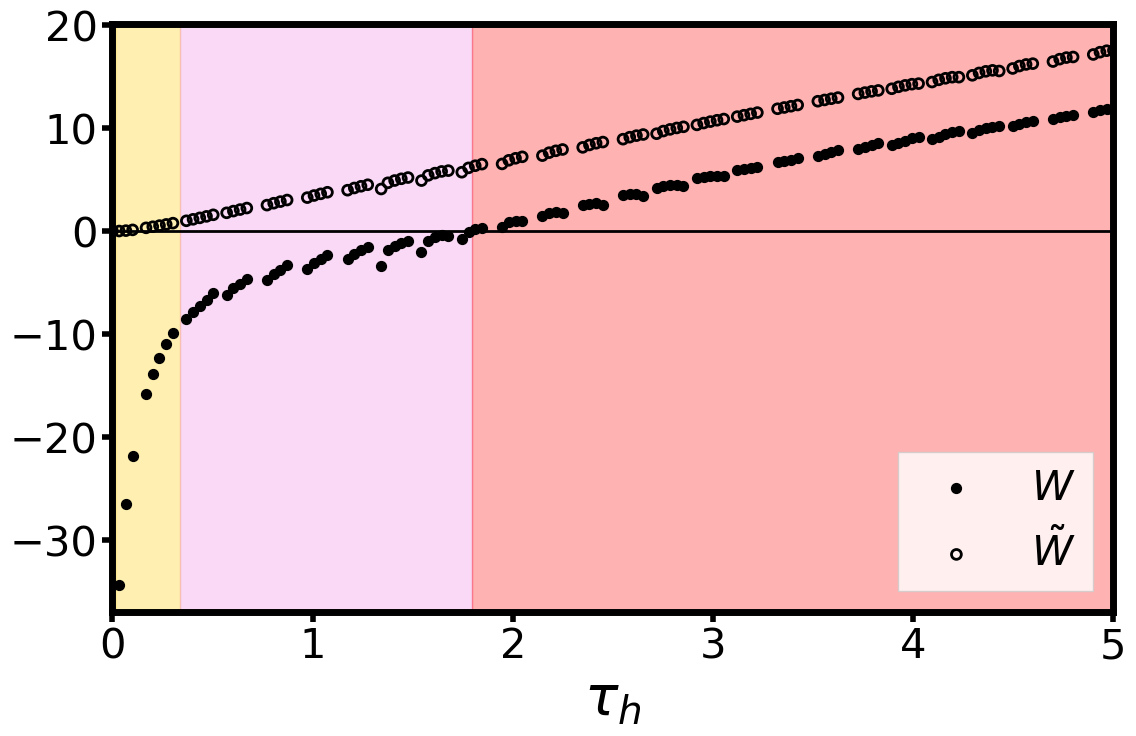}
    \caption{The work output $W$ (filled circles)
    as a function of $\tau_h$ for fixed $\tau_{hc}=2.33$.
     The parameters used are $\omega_c=1,\omega_h=15, T_c=1,T_h=100$
     and $\Omega=10$. Open circles are
    $\widetilde{W}$, which is the  work when the interaction energy is 
    {\it not} included (see Eq.~(\ref{tildew})).
    }
    \label{fig:w}
\end{figure}

\begin{figure}[b]
    \centering
    \includegraphics[width=0.78\linewidth]{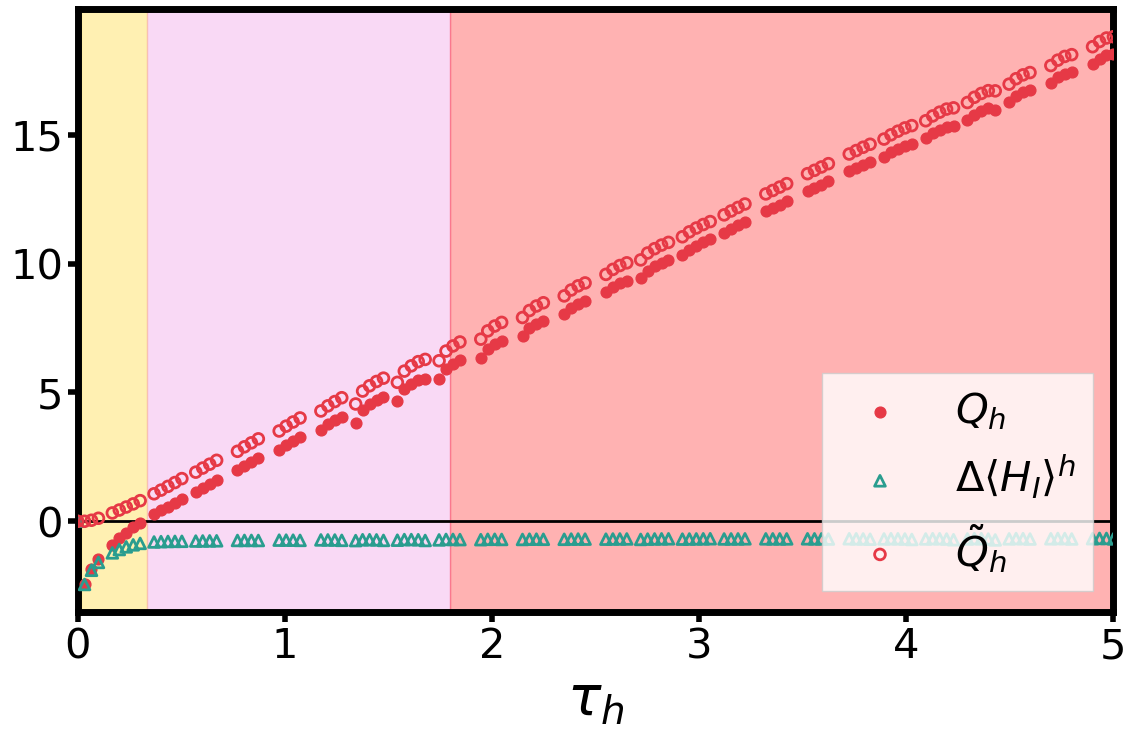}
    \includegraphics[width=0.8\linewidth]{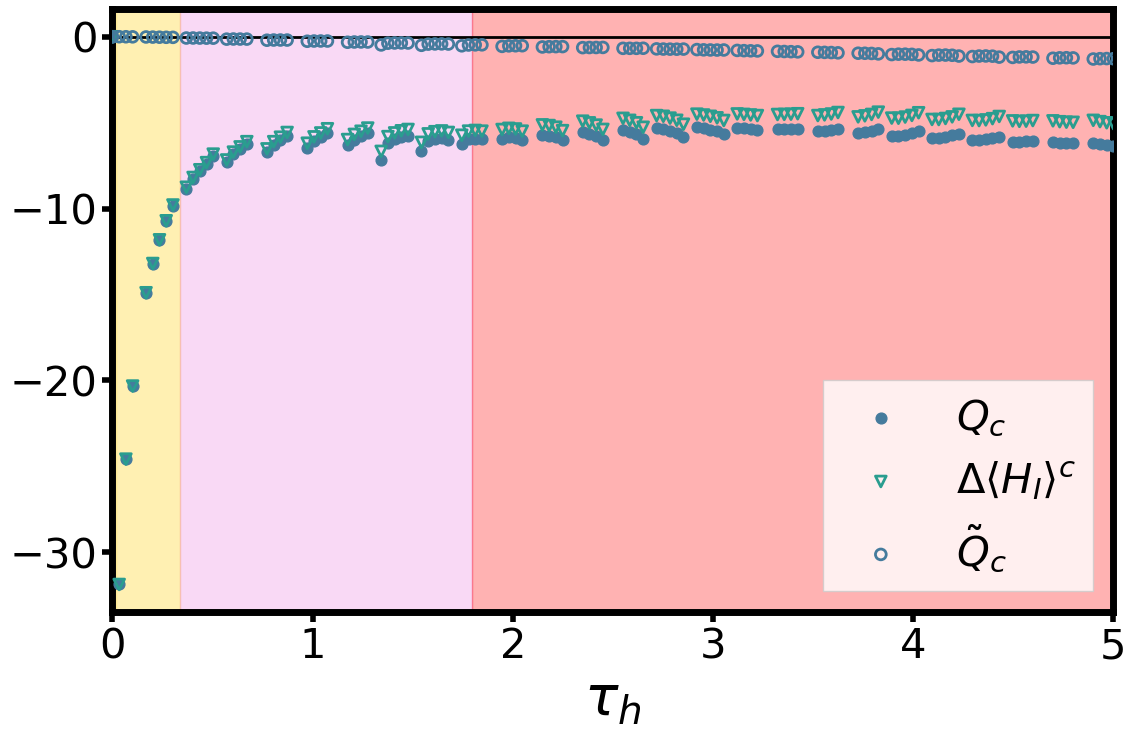}
    \caption{Filled circles are the heat from the hot and cold bath, $Q_h$, $Q_c$, respectively
    as a function of $\tau_h$ for fixed $\tau_{hc}=2.33$ for the case shown
    in Fig.~\ref{fig:w}.
    $\widetilde{Q}_h$ and $\widetilde{PQ}_c$ are the 
    corresponding heat when the interaction energy is {\it not} included (see Eq.~(\ref{tildeq})).
    $\langle \Delta H_I^{(\nu)}\rangle$ (triangles) is the change in the interaction energy
    during the hot $\nu=\mathrm{h}$ and cold $\nu=\mathrm{c}$ isochoric processes (see Eq.~(\ref{DeltaHI})).}
    \label{fig:qh}
\end{figure}

The fact that there is a nonvanishing contribution from the interaction energy has a significant 
implication on the nature of thermodynamics of the
Otto cycle, since the definitions of work and heat include 
$\langle \Delta H^{(\nu)}_I\rangle$  as discussed in Eqs.~(\ref{work_int}), (\ref{qh}) and (\ref{qc}).
In particular, the negative contribution of the 
change in the interaction energy implies that 
the work output will be smaller than that for the case where the 
interaction energy is neglected, as can be seen from Eq.~(\ref{work_int}).
An example of this is shown in Figs.~\ref{fig:w} and \ref{fig:qh}. In that particular case,
we look at how the work $W$ and heats, $Q_h$ and $Q_c$ 
behave depending on the isochoric time $\tau_h$ for 
fixed adiabatic time $\tau_{ch}$.
As we can see in Fig.~\ref{fig:w}, if we consider 
the work $\widetilde{W}$, which is extracted only from the 
system energy change during the adiabatic processes,
as given in Eq.~(\ref{tildew}), it is positive for all steady states with varying $\tau_h$.
However, the change in the interaction energy $\langle \Delta H^{(\nu)}_I\rangle$
stays negative for hot and cold baths, as can be seen in Fig.~\ref{fig:qh}.
Therefore, when the interaction energy is taken into account, 
the work output $W$ becomes negative for short isochoric time $\tau_h$ 
and becomes positive for larger $\tau_h$ as shown in Fig.~\ref{fig:w}. 
This means that the Otto cycle in that
particular setting does not work as an engine for  short isochoric times,
but works as one for longer $\tau_h$. 

The negative contribution of the interaction energy also has an implication on 
the heat intake $Q_h$ from the hot bath. As we can see from the
top panel of Fig.~\ref{fig:qh}, $Q_h$ is negative for short isochoric times, then
changes sign and stays postive for longer $\tau_h$. This means that the Otto cycle
dumps heat into the hot bath for a short isochoric time
$\tau_h$ and begins to absorb heat only for longer $\tau_h$. 
We note that the heat from the cold bath $Q_c$ remains negative for all $\tau_h$,
as can be seen from the lower panel of Fig.~\ref{fig:qh}. That is, the Otto cycle always
dumps heat to the cold bath.

\begin{figure}[b]
    \centering
    \includegraphics[width=0.9\linewidth]{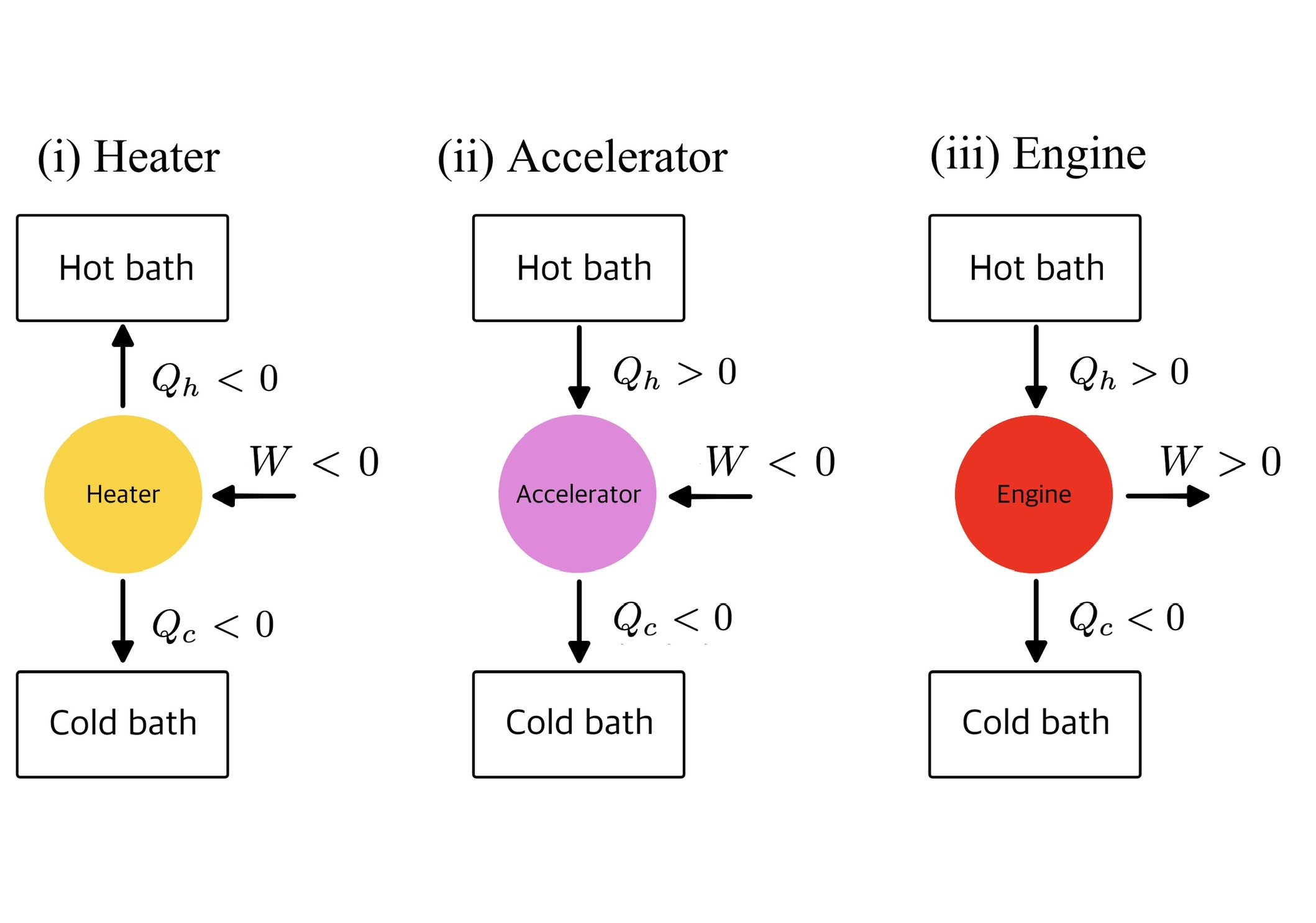}
    \caption{Modes of operation for the Otto cycle}
    \label{fig:cases}
\end{figure}

Combining these facts, we conclude that our Otto cycle can operate in three different modes 
depending on the isochoric and adiabtic times $\tau_h$ and $\tau_{ch}$, respectively.
In fact, through our numerical calculations, we find that, for fixed adiabatic time $\tau_{ch}$, we have $W<0$, $Q_h<0$ for short isochoric times
$\tau_h$ satisfying $0<\tau_h<\tau_h^{(1)}$ (see Fig.~\ref{fig:cases} (i)) for some $\tau^{(1)}_h$.
The Otto cycle uses work from outside to heat both hot and cold baths. We call this 
a heater (H) following Ref.~\cite{gatto2026}. For $\tau_h^{(1)}<\tau_h<\tau_h^{(2)}$ for some $\tau_h^{(2)}$,
we have $W<0$ and $Q_h>0$ (see Fig.~\ref{fig:cases} (ii)). This is referred to
as an accelerator (A). Only for $\tau_h>\tau_h^{(2)}$,
the Otto cycle starts to operate as an engine (E) where $W>0$ and $Q_h>0$ 
(see Fig.~\ref{fig:cases} (iii)). We do not find any cases of a refrigerator for which 
$W<0$, $Q_h<0$ and $Q_c>0$. These three cases are indicated in Figs.~\ref{fig:w} and \ref{fig:qh} 
with different colors.

\begin{figure}[t]
    \centering
    \includegraphics[width=0.9\linewidth]{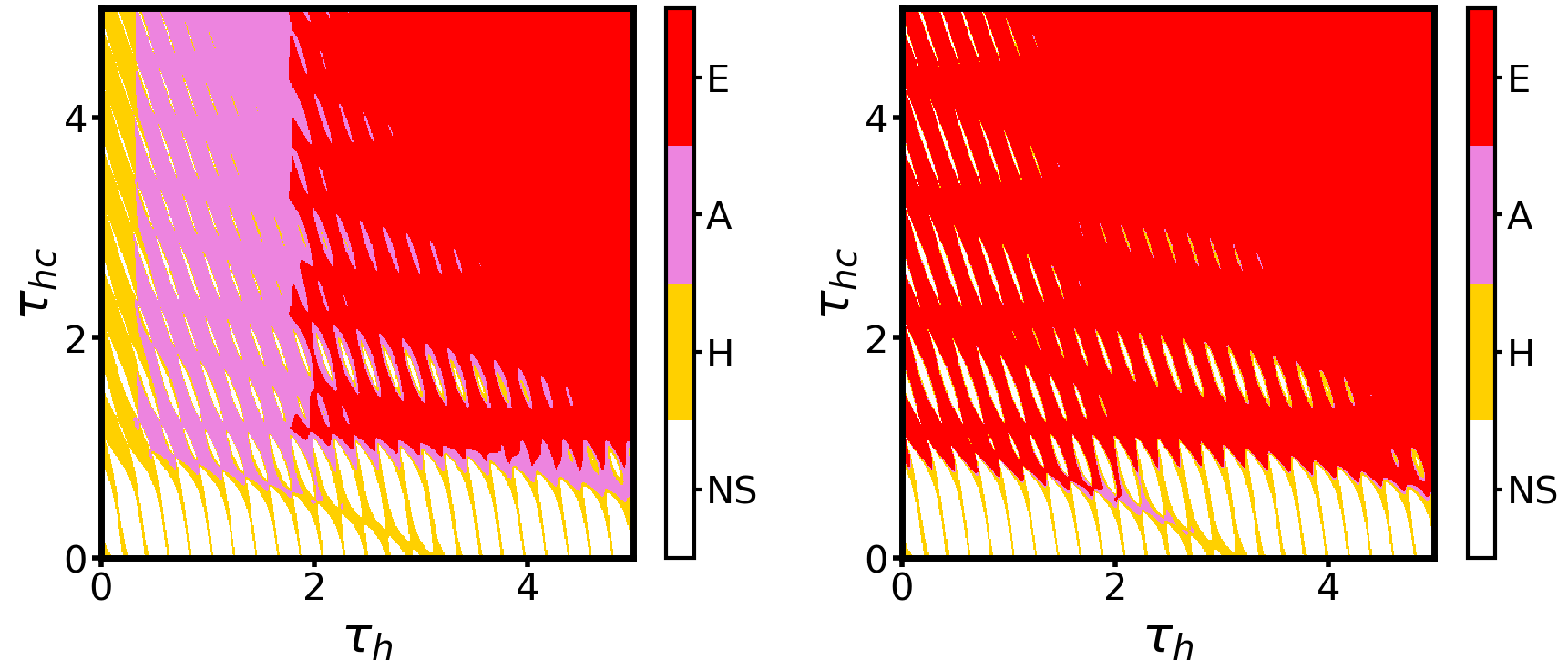}
    \includegraphics[width=0.9\linewidth]{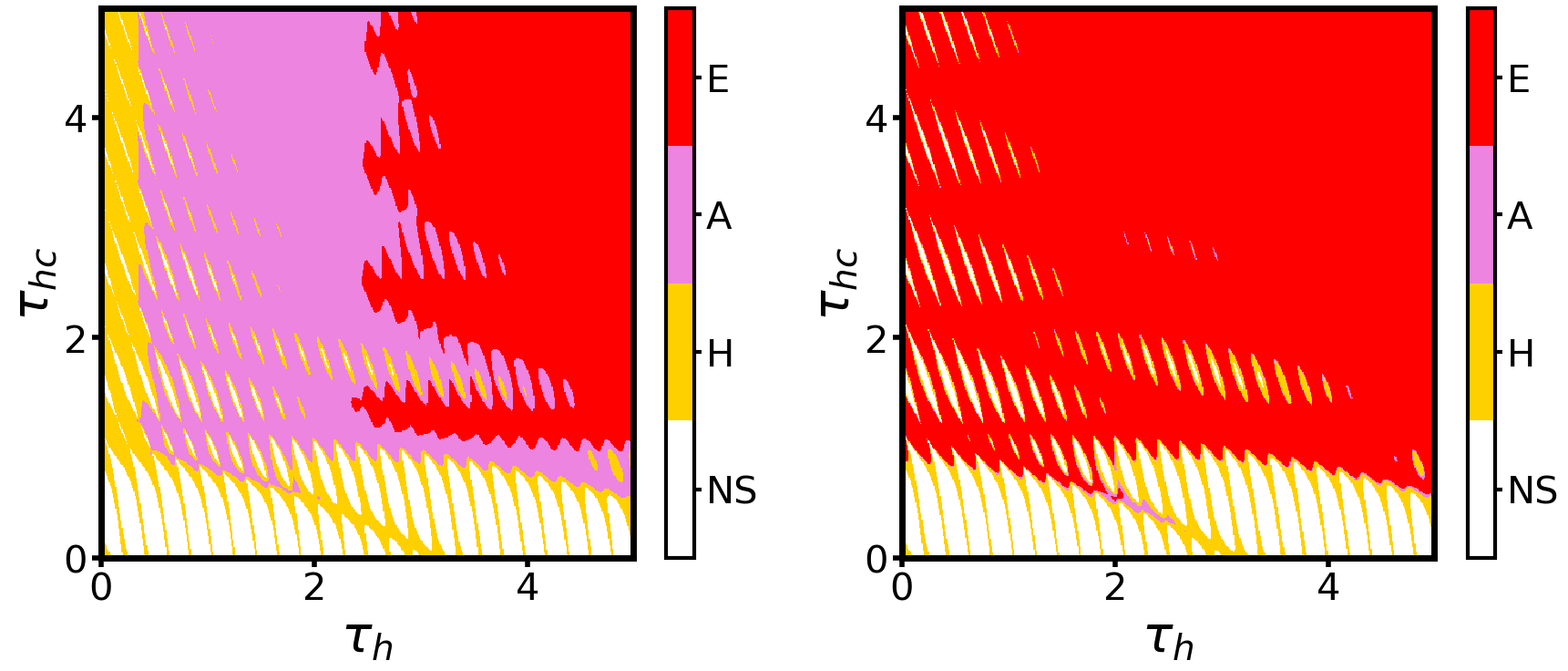}
    \caption{Modes of operations of the Otto cycle for weak coupling $\gamma=0.1$. Three different modes,
     engine (E), accelerator (A) and heater (H), are labeled by 
    different colors. The white regions (NS) are where there is no cyclic steady state. 
    The figures on the left columns are obtained from the exact Heisenberg-Langevin equations, 
    while those on the right column are from the corresponding Markovian approximation
    using the Lindblad equation. 
    The cutoff frequencies are $\Omega=10$ and $20$ for the upper and lower panels, respectively.
    The other parameters used are $\omega_c=1,\omega_h=15, T_c=1,T_h=100$.}
    \label{fig:types_weak}
\end{figure}

\begin{figure}[t]
    \centering
   \includegraphics[width=0.9\linewidth]{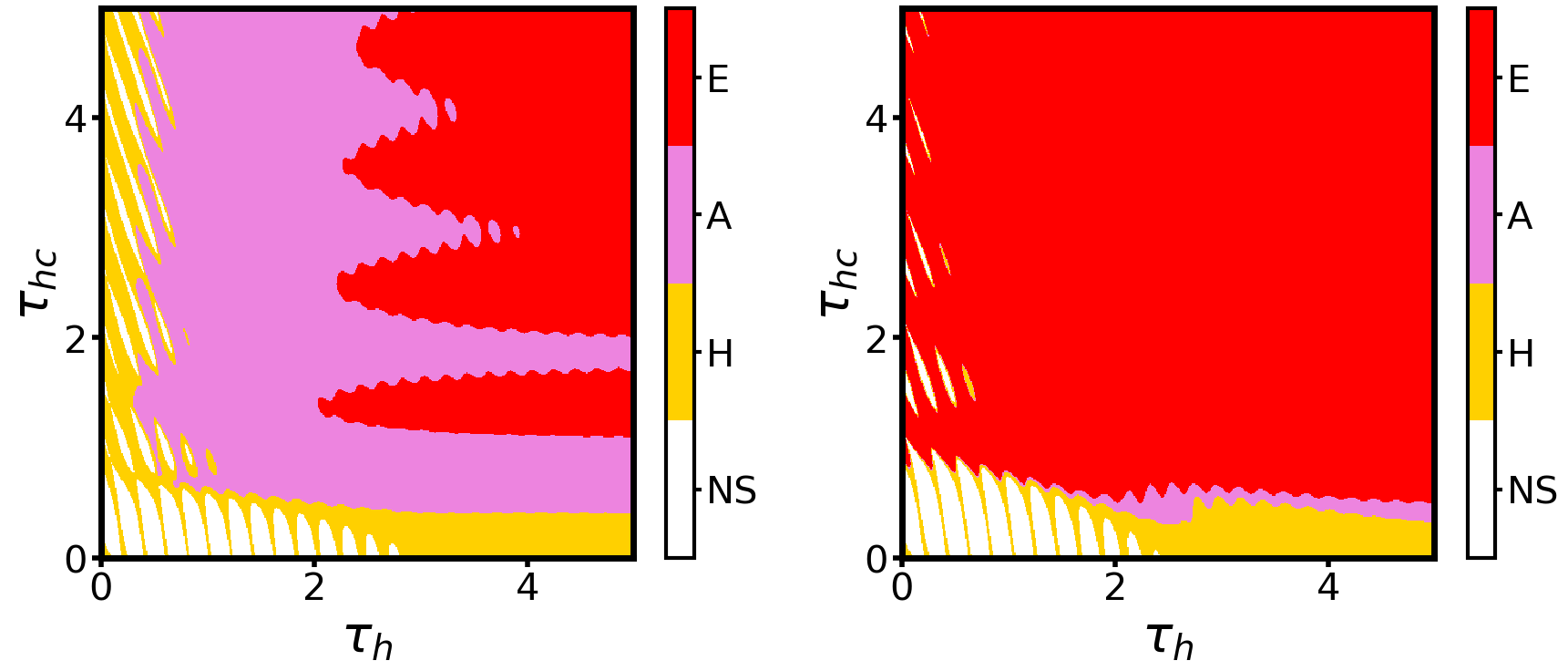}
    \includegraphics[width=0.9\linewidth]{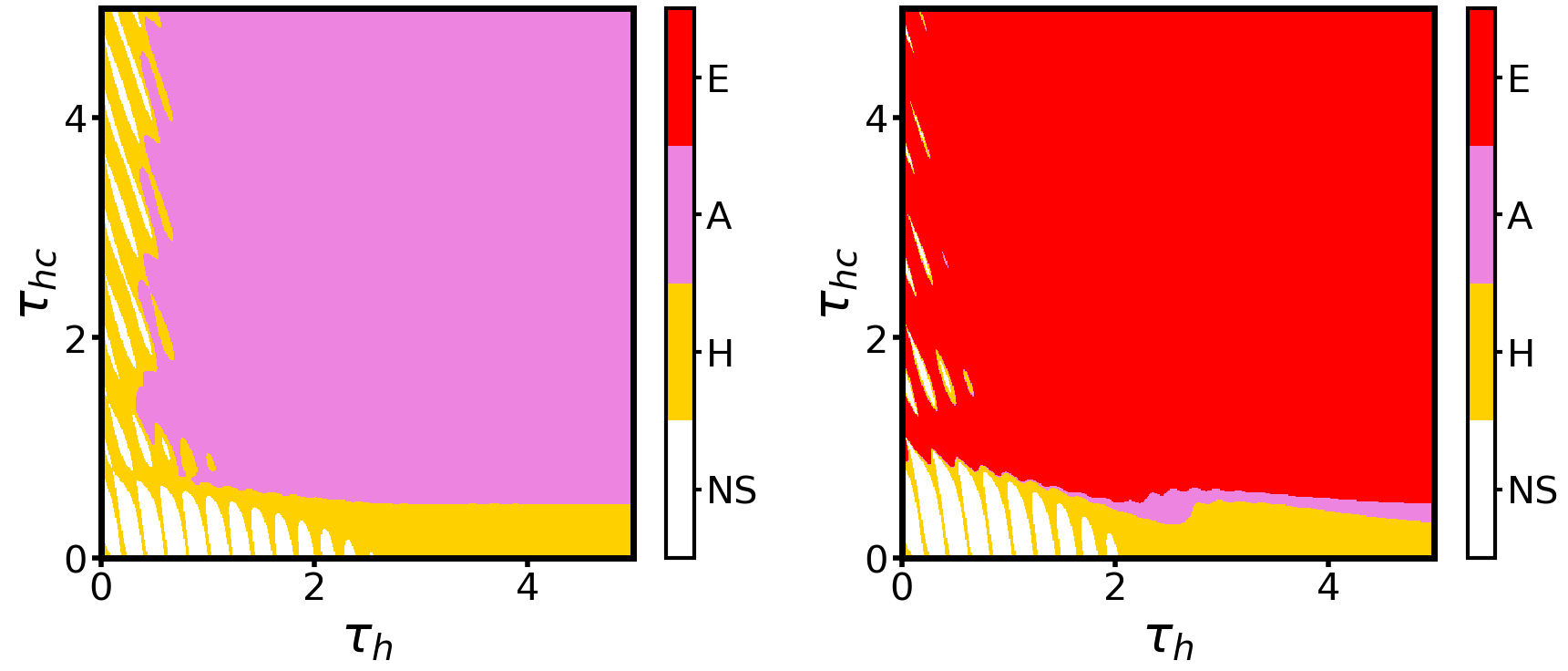}
    \caption{Modes of operations as given in Fig.~\ref{fig:types_weak} but for strong coupling $\gamma=0.7$. The other parameters the same as in Fig.~\ref{fig:types_weak}. }
    \label{fig:types_strong}
\end{figure}

Our finding is in contrast to the results obtained from the Markovian approximation, where
the isochoric process is described by the corresponding Lindblad equation, Eq.~(\ref{lindblad}).
We find that, for the same set of parameters, 
the Otto cycle described by the Markovian approximation 
appears to operate only as an engine. This is illustrated in Figs.~\ref{fig:types_weak}
and \ref{fig:types_strong}, where we show in which modes (H, A or E) the Otto cycle operates in 
the cyclic steady states obtained for various values of 
the isochoric and adiabatic times, $\tau_h$ and $\tau_{ch}$, respectively. 
The left columns are obtained by solving the exact Heisenberg-Langevin equations.
They show the general change of modes (H$\to$A$\to$E) mentioned in the previous 
paragraph for fixed $\tau_{ch}$ and increasing $\tau_h$.
For the right columns, the Markovian Lindblad equation is used for the isochoric processes.
It shows that the only mode the Otto cycle operates in is an engine and
totally misses out the heater and accelerator parts for short isochoric time.
The difference between the exact and the Markovian approximate results 
increases as the coupling strength $\gamma$ between the system and the bath increases,
as can be seen by comparing Figs.~\ref{fig:types_weak} and \ref{fig:types_strong}.
However, we note that the difference is quite large even for 
a very weak coupling value $\gamma=0.1$ in Fig.~\ref{fig:types_weak}. We expect that 
only in the vanishing coupling limit $\gamma\to 0$, we can see that our exact results 
reduce to those in the Markovian approximation.
Therefore, for experimental systems that correspond to much smaller values
of $\gamma$ than those studied in this paper, the Markovian 
description can still be applied. However, for systems described by
increasing values of $\gamma$, non-Markovian features due to the system-bath interaction 
become apparent, and the thermodynamic properties of Otto cycles differ
drastically from those of the Markovian approximation.

It is natural to see that the effect of the interaction between
the system and the bath on thermodynamic quantities such as $W$, $Q_h$
and $Q_c$ increases with 
the coupling strength $\gamma$. 
Interestingly, we find that, for the thermal baths described by 
the Ohmic bath with the Lorentz-Drude cutoff given by Eq.~(\ref{J}), 
the effect of the interaction between the system 
and the bath depends sensitively on the cutoff frequency 
$\Omega$ of the spectral density.
As we can see by comparing Figs.~\ref{fig:types_weak}
and \ref{fig:types_strong}, the effect of interaction increases
with increasing $\Omega$. In fact, for relatively strong coupling $\gamma=0.7$,
the Otto cycle does not operate as an engine within the isochoric time studied in
Fig.~\ref{fig:types_strong}. We expect that the Otto cycle begins to work as an 
engine if we use longer $\tau_h$. This dependence on the cutoff frequency is 
the result of the particular form of the dissipative kernel given in Eq.~(\ref{eta_t}) which is proportional 
to $\gamma\Omega^2$ for the thermal bath with the spectral density
with the Lorentz-Drude cutoff.

Another point we have to mention with regard to Figs.~\ref{fig:types_weak}
and \ref{fig:types_strong} is the existence of the recurring regions where
no cyclic steady state is found. This is indicated in those figures as 
a collection of white lines. The same phenomenon was observed 
in the study of the Otto cycle in the Markovian approximation \cite{lee2020,Insinga_2018}. 
It has been attributed to resonances that occur when the frequency difference
$\omega_h-\omega_c$ and the operation times $\tau_h$ and $\tau_{ch}$ satisfy 
a certain condition \cite{lee2020}. We believe that in our exact treatment
the loss of steady states is also caused by the same mechanism.

\begin{figure}[t]
    \centering
    \includegraphics[width=0.9\linewidth]{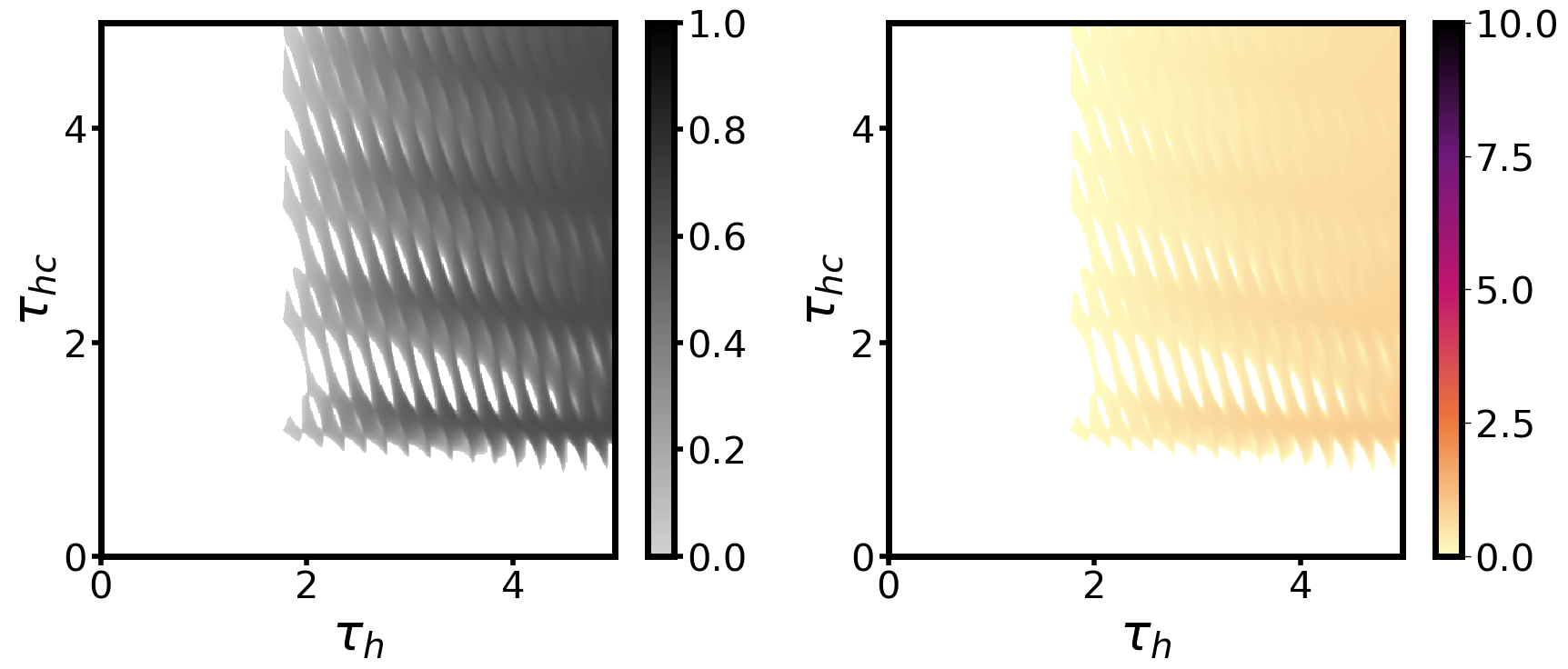}
    \includegraphics[width=0.9\linewidth]{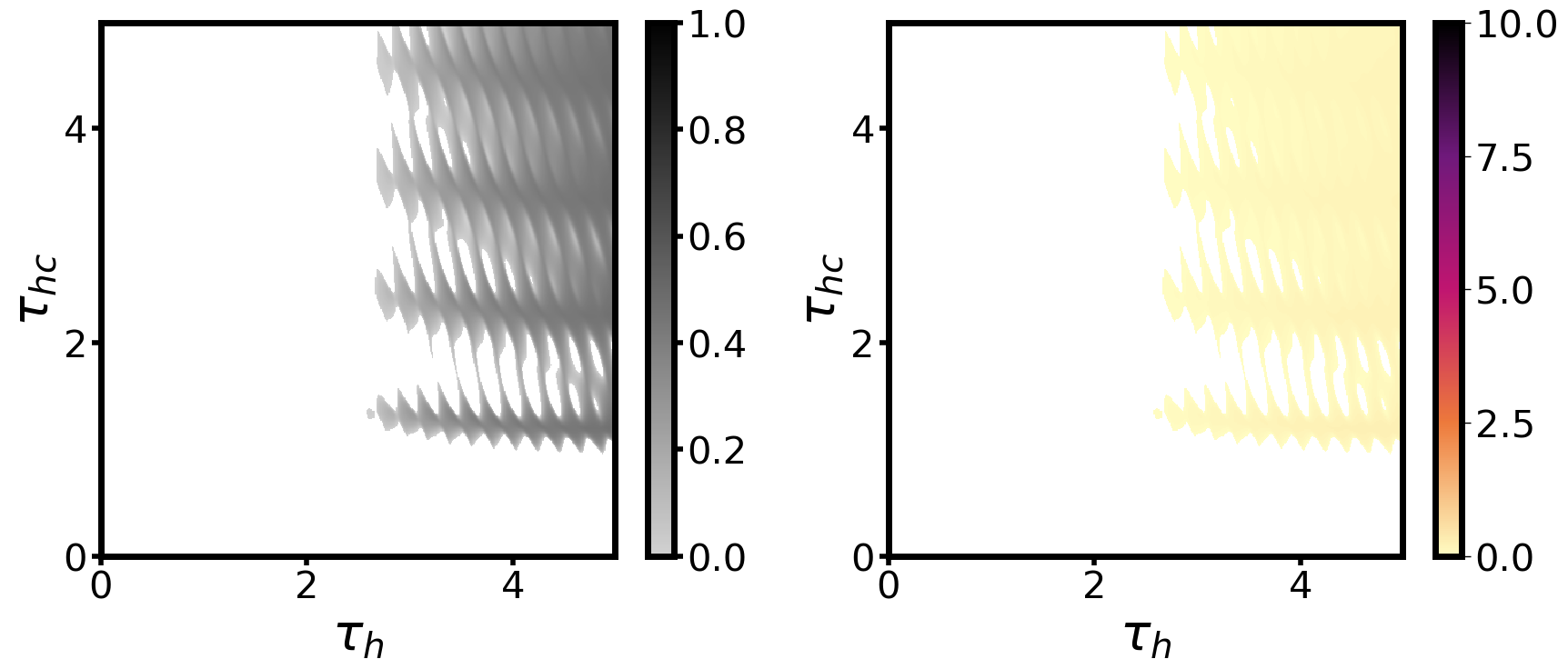}
\caption{Efficiency $\eta$ (left column) and the power $P$ (right column) of the Otto cycle
when it operates as an engine as a function of adiabatic and isochoric times, $\tau_{hc}$ and $\tau_h$, respectively. The parameters used are $\omega_c=1,\omega_h=15,\Omega=10,T_c=1,\gamma=0.1$
The difference between the upper and lower panels are the temperature of the hot bath: $T_h=100$ (upper) and $T_h=50$ (lower). }
    \label{fig:effpower}
\end{figure}

When the Otto cycle operates as an engine, we can define the 
efficiency $\eta$ and the power $P$ as
\begin{align}
    \eta=\frac{W}{Q_h},~~
    P=\frac{W}{\tau_{\rm cyc}}.
\end{align}
Examples of the efficiency and power of the Otto engine as a function of adiabatic and isochoric times are shown in Fig.~\ref{fig:effpower}. We can see that for fixed $\tau_{hc}$ the efficiency increases slowly as $\tau_h$ increases. We expect that in the long isochoric time limit $\eta$ approaches the Otto efficiency $\eta_{\rm Otto}=1- \omega_c/\omega_h$. This is in contrast to 
the correspon ding Markovian case. We find that the efficiency for the Markovian Otto engine quickly approaches $\eta_{\rm Otto}$
even for small values of $\tau_{hc}$ and $\tau_h$. 

\begin{figure}[t]
    \centering
     \includegraphics[width=0.7\linewidth]{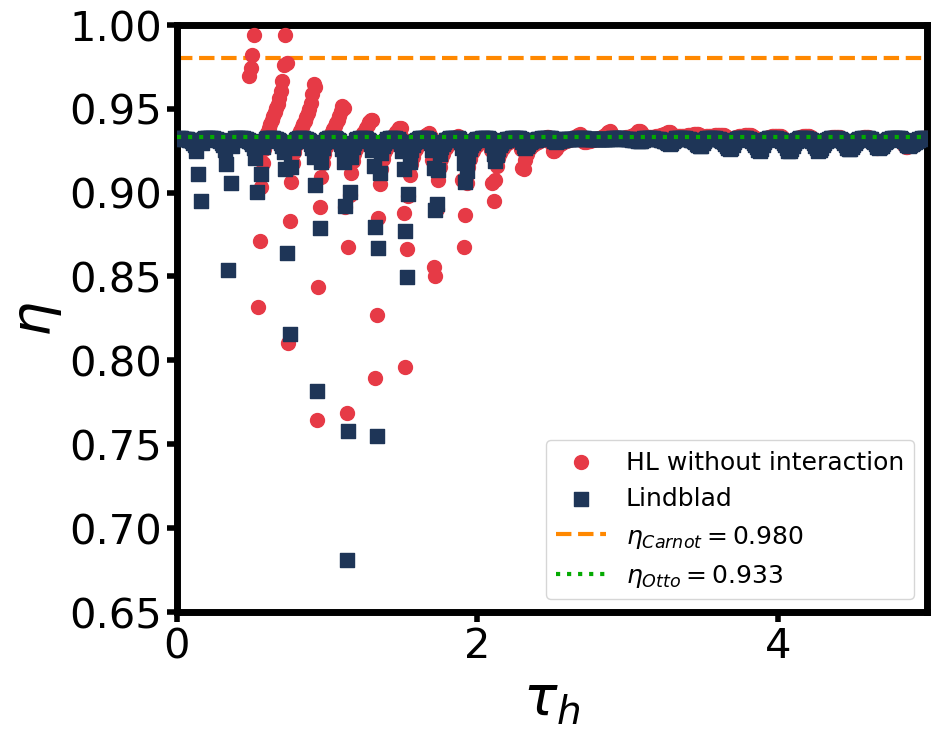}
    \caption{The efficiencies calculated from Heisenberg-Langevin (HL) 
    equations without interaction (circles) and from the 
    Lindblad equation in the Markovian approximation (squares)
    as functions of the isochoric time $\tau_h$ for fixed adiabatic time $\tau_{hc}=3.4$. 
    The Carnot and Otto efficincies, $\eta_{\rm Carnot}$ and $\eta_{\rm Otto}$, respectively are indicated as horizontal
    lines. The parameters used are $\omega_c=1,\omega_h=15,\Omega=10,T_c=1, T_h=50$ 
    and $\gamma=0.1$.}
    \label{fig:eff}
\end{figure}

We note that if we somehow ignore the effect of interaction in our Heisenberg-Langevin approach and calculate
the efficiency and power from $\widetilde{W}$ and $\widetilde{Q}_h$ in Eqs.~(\ref{tildew})
and (\ref{tildeq}) alone, the results are quite similar to those obtained from the Markovian 
approximation. This is in some ways expected, since the changes in the system variables of 
the Heisenberg-Langevin and Lindblad approaches
are quite similar to each other 
as we can see in the upper panels of Figs.~\ref{fig:evolution1} and \ref{fig:evolution2}.
However, there is a crucial difference between them. If we use $\widetilde{W}$ and
$\widetilde{Q}_h$ and neglect the effect of interaction to calculate the efficiency, we
find that this efficiency sometimes gives values higher than the Carnot efficiency
$\eta_{\rm Carnot}=1-T_c/T_h$. This is illustrated in Fig.~\ref{fig:eff}.
This point has been raised in previous studies of non-Markovian Otto engines \cite{Zhang_2014}.
It demonstrates that including the effect of interaction is crucial in studying 
non-Markovian Otto cycles.

The efficiency and power of Markovian Otto engines have been studied extensively.
One notable result among these is the one that deals with the so-called trade-off
relation between efficiency and power for the Markovian quantum Otto engine
\cite{Chun_Park_2025}. For classical thermal engines, there have
been extensive studies on the power-efficiency
trade-off relations \cite{Shiraishi_Saito_Tasaki_2016,Dechant_Sasa_2018,Pietzonka_Seifert_2018}. 
In Ref.~\cite{Chun_Park_2025}, using the phase space approach 
in analyzing the Lindblad equation, the authors
applied the technique developed in classical stochastic thermodynamics 
to the time evolution equation for the quasiprobability distribution
and derived the universal inequality for efficiency $\eta$ and power $P$ 
for the Markovian quantum Otto engine \cite{Chun_Park_2025}. 
Their results are for the case where the coupling constant
$\gamma_L$ in the Lindblad equation Eq.~(\ref{lindblad}) is the same for the
isochoric processes in contact with the hot and cold baths. In our case, however,
the Markovian limit of the exact time evolution equation results in two different 
coupling constants $\gamma_L^{\rm (h)}$ and $\gamma_L^{\rm (c)}$ as we have shown in Eq.~(\ref{lindblad}). The power-efficiency trade-off relation derived in 
Ref.~\cite{Chun_Park_2025} can be generalized to the present case and we have 
\begin{align}
    \frac{P}{P_0}\leq \frac{\eta(\tilde{\eta}-\eta)}{1-\eta},
    \label{tradeoff}
\end{align}
where
\begin{align}
    \tilde \eta=1-\frac{\tilde T_c}{\tilde T_h}
\end{align}
with the effective temperatures $\tilde{T}_\nu$ $(\nu=\mathrm{h,c})$ given by
\begin{align}
    \tilde{T}_\nu =&\omega_\nu \left(\bar n_\nu+\frac{1}{2}\right) = 
    \frac{\omega_\nu}{2}\coth\left(\frac{\omega_\nu}{2 T_\nu}\right).
\end{align}
In Eq.~(\ref{tradeoff}), $P_0$ is given by
\begin{align}
    P_0 \equiv \tilde T_h \phi_\text{min},
    \label{P0}
\end{align}
where
\begin{align}
    \phi_\text{min}=\text{min}\left(\gamma^{(h)}_L\frac{\tau_{h}}{\tau_{cyc}},\gamma^{(c)}_L\frac{\tau_{c}}{\tau_{cyc}}\right) .
\end{align}
Note that the Markovian approximation of our model yields two different
coupling constants $\gamma^{(\nu)}_{L}$ $(\nu=\mathrm{h,c})$ as
given by Eq.~(\ref{eff_gamma}).

\begin{figure}[b]
    \centering
    \includegraphics[width=0.7\linewidth]{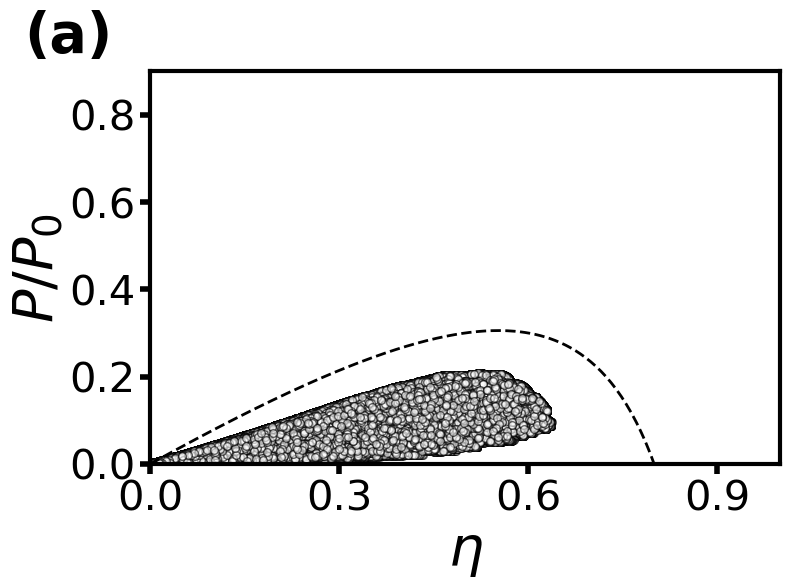}
    \includegraphics[width=0.7\linewidth]{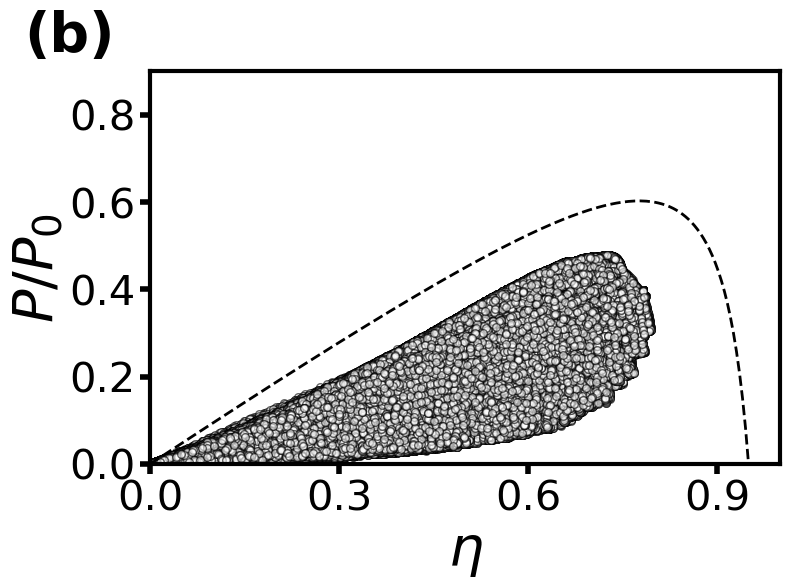}
    \caption{Collection of approximately $10^7$ data points of the efficiency $\eta$
    vs.\ power $P$ normalized by $P_0$ in Eq.~(\ref{P0}). The data points are
    generated by varying all the parameters of the present non-Markovian Otto 
    engine for fixed (a) $\tilde{\eta}=0.8$
    and (b) $\tilde{\eta}=0.95$. 
    The dashed lines are the right hand side of Eq.~(\ref{tradeoff}), which is
    the upper bound of the Markovian power-efficiency trade-off relation.
    }
    \label{fig:tradeoff}
\end{figure}

In Fig.~\ref{fig:tradeoff}, we plot our results for the efficiency and power obtained
from exact Heisenberg-Langevin equations against
this Markovian trade-off bound (the right hand side of Eq.~(\ref{tradeoff})).
For two fixed values of $\tilde{\eta}$, we vary the other parameters and collect the cases
where the Otto cycle works as an engine and calculate $\eta$ and $P$ for each case.
In both values of $\tilde{\eta}$, we find that our non-Markovian Otto engine stays well below this 
Markovian maximum power line given by the trade-off relation.

\section{Discussion and Summary}

We have studied a non-Markovian quantum Otto cycle consisting of a harmonic oscillator 
by solving exact Heisenberg-Langevin equations. We were able to obtain analytic expressions
for the isochoric and adiabatic propagators, which allowed us to explore  
the parameter space of the Otto cycle exhaustively and find cyclic steady states without much 
numerical cost. We were also able to find analytic expressions for the interaction 
energy between the system and the bath. We find that the inclusion of the interaction 
energy in heat and work has a dramatic effect
on the property of the Otto cycle even for small coupling strength between the system and the bath. Indeed, if we only look at the system variables,
there is little difference between the results obtained from the Heisenberg-Langevin equations and those from the Markovian approximation for small coupling strength. However, when the interaction energy is properly incorporated into heat and work, we can see that
the non-Markovian Otto cycle behaves very differently 
from  the Markovian counterpart. 

Depending on the isochoric time, our Otto cycle operates
in three different modes, as shown in Sec.~\ref{sec:results}. 
A similar behavior was observed in Ref.~\cite{Ishizaki_2023} for a qubit system.
For short isochoric time $\tau_h$, the Otto cycle behaves as a heater and then
an accelerator as $\tau_h$ increases, where the system uses external work to heat the baths.
These transient behaviors are absent in the Markovian counterpart. 
For longer $\tau_h$, the Otto cycle performs as an engine.
In this paper, we use a sharp step-function protocol for the attachment and detachment of the system and the bath.
If we use a different protocol as, for example, in Ref.~\cite{Wiedmann_2020}, we still
expect that similar behavior occurs with some quantitative changes in the shape of the region
of these modes in the $\tau_h$-$\tau_{hc}$-plot like Figs~\ref{fig:types_weak}
and \ref{fig:types_strong}. It would be interesting to see how the transient behaviors 
depend on the attachment/detachment mechanism.
However, an investigation along this line  
would be a challenging task, as it requires a numerical integration of
the Heisenberg equations, and it would prevent us from doing a comprehensive study 
on the parameter space, as we have done in the paper using analytical solutions.
The transient behaviors may also be influenced by the preparation of the initial state
of the isochoric process. In this paper, we used the product initial state between the system and the bath.
A quantitative analysis on the behavior of Otto cycles for 
initial states other than the product state
would have the same numerical difficulty. We leave these lines of investigation for future study.

Non-Markovianity is usually associated with the presence of
memory effects, but quantifying the effect of non-Markovianity in open quantum systems
is a nontrivial task and has been the subject of intensive study 
\cite{breuer_colloquium_2016, rivas_quantum_2014}. 
In Ref.~\cite{einsiedler_non-markovianity_2020}, the non-Markovianity of 
the quantum Brownian motion
described by the Caldeira-Leggett model has been studied quantitatively 
using the non-Markovianity measure 
proposed in Ref.~\cite{breuer_measure_2009}. The results show a nontrivial and non-monotonic 
dependence of the non-Markovianity measure on the coupling constant $\gamma$, 
the cutoff frequency $\Omega$ and the temperature $T$. In this paper,
we have only looked at a limited range of these parameters. It would be interesting 
to see more explicitly how the non-markovianity measure is related to the thermodynamic 
properties of the Otto cycles that we found in this paper.

In this paper, we have used a special protocol Eq.~(\ref{protocol}) 
of the frequency change for adiabatic processes. It has been widely
used \cite{Kosloff_Rezek_2017, lee2020} for Markovian Otto engines. 
The protocol itself has no physical significance except 
for the analytical tractability of the adiabatic propagator Eq.~(\ref{padia}).
We could of course incorporate a more general protocol for the frequency change into our
calculation. This would involve numerical evaluations of the adiabatic propagator for
each cycle and would make the calculation more complicated. However, we believe that 
any other protocol does not make significant qualitative changes to our results.
We believe that even with a general protocol 
the same change of modes of operation as we have seen in Sec.~\ref{sec:results} occurs, namely the one from dumping heat into the bath to extracting work as an engine 
as the isochoric time increases.

An interesting area of possible future research is the dependence
of the quantum Otto cycle on
the spectral density of the bath. In this paper, we used the Ohmic
spectral density with the Lorentz-Drude cutoff. In Ref.~\cite{Picatoste_2024},
a Lorentzian spectral density with a peak at some frequency was considered.
It was shown there that the Otto engine can have an enhanced work output 
compared to the Markovian one, which seems to be in contrast to our results. 
We note that in Ref.~\cite{Picatoste_2024}, the effect of the interaction
was not considered as in the present paper. It would be interesting to see whether
this kind of spectral density makes a qualitative difference within our scheme.

It is interesting to see that
our power-efficiency data for the non-Markovian Otto engine 
fall below the Markovian bound. 
This results from the fact that within our model
the effect of the interaction is to reduce the work 
output of the non-Markovian engine compared to the Markovian counterpart. 
As mentioned above, we have only looked
at one specific protocol for adiabatic processes. Therefore, we cannot claim conclusively
that this is always true for all non-Markovian thermal engines. 
It remains to be seen 
whether this result holds for all general non-Markovian engines,
as the enhancement of engine performance has been seen in some non-Markovian systems \cite{Ptaszy_2022}.
We are not aware of 
any theoretical attempt to obtain a non-Markovian power-efficiency bound. 
It would be interesting to see whether such a bound for non-Markovian 
quantum engines can be found and, if it exists, it stays below the Markovian one.

\begin{acknowledgments}
This work was supported by NRF
grant funded by the Korea government (MSIT) (RS-2023-00276248).
We thank Hyun-Myung Chun and Jong-Min Park for many useful comments and discussions, especially
on the power efficiency trade-off relation for the Markovian Otto engine.
\end{acknowledgments}

\appendix
\section{The Isochoric Process}
\label{app:1}
Here we sketch the calculations leading up to the isochoric propagator
given in Sec.~\ref{sec:isochore} for the Lorentz-Drude spectral density
in Eq.~(\ref{J}). 
Taking the Laplace transform
of Eq.~{\ref{EOM}} using 
\begin{align}
    \hat{x}(s)=\int_0^\infty dt\; e^{-st} x(t)
\end{align} 
and back transform it,
we can easily show that the solution is given by
Eq.~(\ref{xsol}) with
\begin{align}
\hat{G}_{2}(s)=\frac{\hat{G}_{1}(s)}{s}=\frac{1}{s^{2}+s \hat{\gamma}(s)+\omega^{2}_\nu}, \label{green}
\end{align}
where the Laplace transform of $\gamma(t)$ in Eq.~(\ref{gamma}) for the spectral 
density in Eq.~(\ref{J}) is given by
\begin{align}
\hat{\gamma}(s)=2 \gamma \frac{\Omega}{s+\Omega} . \label{gamma:lap}
\end{align}
Note that, for simplicity of the notation, we  drop the superscript
$(\nu)$ in this Appendix. For both hot ($\nu=\mathrm{h}$) or
cold ($\nu=\mathrm{c}$) isochores, the calculations are the same except for the 
appearance of $\omega_\nu$ for the corresponding bath $\nu$.
Note also that Eq.~(\ref{green}) indicates that $G_1(t)=
\dot{G}_2(t)$.
Combining Eqs.~(\ref{green}) and (\ref{gamma:lap}), we have
\begin{align}
\hat G_2(s)=&\frac{s+\Omega}{s^3+\Omega s^2 +(2\gamma \Omega+ \omega^2_\nu )s +\omega^2_\nu \Omega} \label{hatG2} \\
=& \sum_{i=1}^3 \frac{d_i}{s-s_i}, \label{hatG2_1}
\end{align}
where $s_i$ are three solutions to the cubic equation for $s$ obtained by setting the denominator of Eq.~(\ref{hatG2}) to zero and 
\begin{align}
d_{i}=\frac{s_{i}+\Omega}{\prod_{j \neq i} ( s_{j}-s_{i}) } .\label{d_i}
\end{align}
We therefore have
\begin{align}
 G_2(t)&=\sum_{i=1}^3 d_i e^{s_i t} \label{G_2}
\end{align}
and
\begin{align}
 G_1(t)&=\sum_{i=1}^3 d^\prime_i e^{s_i t}, \label{G_1}
\end{align}
where $d^\prime_1=s_i d_i$.

Now, from Eq.~(\ref{xsol}) and $p(t)=m\dot{x}(t)$, we have
\begin{align}
\langle x^2(t)\rangle = &G_{1}^2(t)\langle x^2(0)\rangle + \frac{1}{m^2}G_{2}^2(t)\langle p^2(0)\rangle\notag
\\ +& \frac{1}{m}G_{1}(t)G_{2}(t)\langle \{x(0),p(0)\}\rangle + \langle I_{xx}(t) \rangle ,\label{sigma_xx}
\end{align}
\begin{align}
\langle p^2(t)\rangle=& m^2 \dot G_1^2(t)\langle x^2(0)\rangle+\dot G_2^2(t)\langle p^2(0)\rangle\notag
\\+& m\dot G_1(t)\dot G_2(t)\langle \{x(0),p(0)\}\rangle+\langle I_{pp}(t)\rangle \label{sigma_pp}
\end{align}
and
\begin{align}
&\frac 1 2 \langle \{x(t),p(t)\}\rangle \label{sigma_xp}\\
=&  
m\dot G_1(t)G_1(t)\langle x^2(0)\rangle
+\frac{1}{m}\dot G_2(t)G_2(t)\langle p^2(0)\rangle
\notag
\\
+&\frac 1 2 (\dot G_1(t)G_2(t)+\dot G_2(t)G_1(t))
\langle \{x(0),p(0)\}\rangle+\langle I_{px}(t)\rangle ,\notag
\end{align}
where $\langle I_{xx}(t)\rangle $ and $\langle I_{pp}(t)\rangle $ are given by Eqs.~(\ref{Ixx}) and (\ref{Ipp}). 
We also have 
\begin{align}
    \langle I_{px}(t)\rangle = \frac 1 {\omega_\nu} \langle I_C(t) \rangle  ,
\end{align}
where $\langle I_C(t) \rangle $ is given in Eq.~(\ref{IC}). Now using the definitions
Eqs.~(\ref{Ht:def}), (\ref{Lt:def}) and (\ref{Ct:def}), we can easily derive the relations
Eqs.~(\ref{Ht}), (\ref{Lt}) and (\ref{Ct}). We can evaluate these quantities for 
the Lorentz-Drude spectral density given in Eq.~(\ref{J}) by noting that 
the bath correlation function
in Eq.~(\ref{D_1}) in this case is given by \cite{OQS_Breuer}
\begin{align}
    D_1(t)=4m\gamma T\Omega^2 \sum_{n=-\infty}^\infty 
    \frac{\Omega e^{-\Omega |t| } -|v_n|e^{-|v_n||t|}}{\Omega^2-v^2_n},
    \label{D_1:exp}
\end{align}
where $v_n =2\pi n T$. By performing the double integrals in Eqs.~(\ref{Ixx}), (\ref{Ipp}),
and (\ref{IC}) explicitly, we obtain 
\begin{align}
\left\langle I_{xx}(t)\right\rangle=&\frac{2 \gamma k_{B} T \Omega^{2}}{m} \sum_{n=-\infty}^{\infty} \frac{1}{\Omega^{2}-v_{n}^{2}} \mathcal{I}_{xx}(t, n) ,
\\
\left\langle I_{p p}(t)\right\rangle =&2m\gamma k_{B} T \Omega^{2}  \sum_{n=-\infty}^{\infty} \frac{1}{\Omega^{2}-v_{n}^{2}} \mathcal{I}_{pp}(t, n) ,
\\
\langle I_{px}(t)\rangle =&2\gamma k_BT\Omega^2\sum_{-\infty}^{\infty}\frac{1}{\Omega^2-v_n^2}\mathcal{I}_{px}(t,n) ,
\end{align}
where
\begin{widetext}
\begin{align}
\mathcal{I}_{xx}(t, n)= & \sum_{i, j=1}^{3} d_{i} d_{j}\left\{\frac{\left|v_{n}\right|}{\left|v_{n}\right|-s_{i}} \right.
\left(\frac{1-e^{\left(s_{i}+s_{j}\right) t}}{s_{i}+s_{j}}-\frac{e^{\left(-\left|v_{n}\right|+s_{i}\right) t}-e^{\left(s_{i}+s_{j}\right) t}}{\left|v_{n}\right|+s_{j}}\right) \notag \\&
-\frac{\Omega}{\Omega-s_{i}}\left(\frac{1-e^{\left(s_{i}+s_{j}\right) t}}{s_{i}+s_{j}}-\frac{e^{\left(-\Omega+s_{i}\right) t}-e^{\left(s_{i}+s_{j}\right) t}}{\Omega+s_{j}}\right)
+\frac{\left|v_{n}\right|}{\left|v_{n}\right|+s_{i}}\left(\frac{1-e^{\left(s_{j}-\left|v_{n}\right|\right) t}}{\left|v_{n}\right|-s_{j}}+\frac{1-e^{\left(s_{i}+s_{j}\right) t}}{s_{i}+s_{j}}\right)\notag 
\\&
\left. -\frac{\Omega}{\Omega+s_{i}}\left(\frac{1-e^{\left(s_{j}-\Omega\right) t}}{\Omega-s_{j}}+\frac{1-e^{\left(s_{i}+s_{j}\right) t}}{s_{i}+s_{j}}\right)\right\} .
\end{align}
We can just replace $d_i d_j$ by $d'_i d'_j$ for $\mathcal{I}_{pp}(t, n)$
and by $d_i d'_j$ for $\mathcal{I}_{px}(t, n)$.

Finally we can summarize the above results into the following isochoric propagator matrix:
\begin{align}
\mathcal{P}^{(\nu)}_{\rm iso}(t) &= \begin{pmatrix}
\frac{1}{2}\left(\frac{1}{\omega_\nu^2}\dot{G}_1^2 + 2G_1^2 + \omega_\nu^2 G_2^2\right) & 
\frac{1}{2}\left(-\frac{1}{\omega_\nu^2}\dot{G}_1^2 + \omega_\nu^2 G_2^2\right) & 
\frac{1}{\omega_\nu}(\dot{G}_1\dot{G}_2 + \omega_\nu^2 G_1 G_2) & 
\langle I_H(t) \rangle \\
\frac{1}{2}\left(\frac{1}{\omega_\nu^2}\dot{G}_1^2 - \omega_\nu^2 G_2^2\right) & 
\frac{1}{2}\left(-\frac{1}{\omega_\nu^2}\dot{G}_1^2 + 2{G}_1^2 - \omega_\nu^2 G_2^2\right) & 
\frac{1}{\omega_\nu}(\dot{G}_1\dot{G}_2 - \omega_\nu^2 G_1 G_2) & 
\langle I_L(t) \rangle \\
\omega_\nu\left(\frac{1}{\omega_\nu^2}G_1\dot{G}_1 + G_2\dot{G}_2\right) & 
\omega_\nu\left(-\frac{1}{\omega_\nu^2}G_1\dot{G}_1 + G_2\dot{G}_2\right) & 
\dot{G}_1 G_2 + \dot{G}_2G_1 & 
\langle I_C(t) \rangle  \\
0 & 0 & 0 & 1
\end{pmatrix}.
\label{Piso}
\end{align}
\end{widetext}

\section{The Adiabatic Process}
\label{app:2}

The Heisenberg equation (\ref{matrixeq}) for constant $\mu$ can be solved by 
changing the variable from $t$ to 
\begin{align}
    \theta(t)\equiv \int_0^t \omega(t')dt' .
    \label{thetat}
\end{align}
We then have
\begin{align}
     \frac{d}{d\theta} \begin{pmatrix}
  H(\theta)\\
  L(\theta)\\
 C(\theta) 
\end{pmatrix}=\underbrace{\begin{pmatrix}
\mu & -\mu & 0 \\
-\mu & \mu & -2  \\
0 & 2 & \mu  \\
\end{pmatrix}}_{\equiv \mathbb{M}}\begin{pmatrix}
  H(\theta)\\
  L(\theta)\\
 C(\theta) 
\end{pmatrix}.
\label{matrixeq2}
\end{align}
The solution is given by
\begin{align}
    \begin{pmatrix}
  H(\theta)\\
  L(\theta)\\
 C(\theta) 
\end{pmatrix}=e^{ \mathbb{M}\theta}\begin{pmatrix}
  H(0)\\
  L(0)\\
 C(0) 
\end{pmatrix}
\end{align}
By diagonalizing $\mathbb{M}$, we can easily find the adiabatic propagator.
The detailed expression for the process with $\omega(0)=\omega_i$ and $\omega(\tau)=\omega_f$ is given by
\begin{widetext}
\begin{align}
  \mathcal{P}_{\rm ad}(\tau)=
    \begin{pmatrix}
\frac{e^{\mu\theta}}{\Lambda^2}(-4+\mu^2\cosh(\Lambda\theta)) & -\frac{\mu}{\Lambda}e^{\mu\theta}\sinh(\Lambda\theta) & \frac{2\mu}{\Lambda^2}e^{\mu\theta}(\cosh(\Lambda\theta)-1) & 0\\
-\frac{\mu}{\Lambda}e^{\mu\theta}\sinh(\Lambda\theta) &  e^{\mu\theta}\cosh(\Lambda\theta) & -\frac{2e^{\mu\theta}}{\Lambda}\sinh(\Lambda\theta) & 0\\
\frac{2\mu e^{\mu\theta}}{\Lambda^2}(1-\cosh(\Lambda\theta)) & \frac{2}{\Lambda}e^{\mu\theta}\sinh(\Lambda\theta) & \frac{e^{\mu\theta}}{\Lambda^2}(\mu^2-4\cosh(\Lambda\theta)) & 0 \\
0 & 0& 0& 1
\end{pmatrix}  ,
\label{padia}
\end{align}
\end{widetext}
where $\Lambda=\sqrt{\mu^2-4}$.
From Eq.~(\ref{thetat}), we have
\begin{align}
    \theta(\tau)=\frac{1}{\mu}\ln\left(\frac{\omega_f}{\omega_i}\right)
\end{align}
Therefore, for adiabatic 
compression, where $\omega_i=\omega_c$ and $\omega_f=\omega_h$, we have
$e^{\mu\theta}=\omega_h/\omega_c$ and $\tau=\tau_{ch}$. On the other hand, for the expansion
precess $\mathrm{h\to c}$, we have
$e^{\mu\theta}=\omega_c/\omega_h$ and $\tau=\tau_{hc}$.

\section{Calculation of the Average Interaction Energy}
\label{app:3}
We suppose that the system contacts the bath 
at time $t=0$ and detaches from it at $t=\tau>0$ in the isochoric process.
In this Appendix we calculate the average of the interaction Hamiltonian $\langle H^{(\nu)}_{I}(t)\rangle$ at arbitrary time $t$ ($0\leq t \leq\tau$)
and the change in the interaction energy 
$\langle \Delta H^{(\nu)}_{I}\rangle =\langle H^{(\nu)}_{I}(\tau)\rangle
-\langle H^{(\nu)}_{I}(0)\rangle$ during the isochoric process. 
Since the calculations for the hot and cold isochores are essentially the same, 
we drop the index $\nu$ in the following for simplicity of notation. 
We can simply put the index $\nu$ back to the appropriate quantities
in the resulting expressions. 

We start from Eq.~(\ref{HInu_sol})
        \begin{align}
        \langle H_I(t)\rangle 
        &=-\frac 1 2 \left\langle 
        \left\{
        x(t), B(t) \right\}\right\rangle \\
        &+ \frac{m}{2}\int_0^t ds\; \eta(t-s) \left\langle \left\{ x(t), x(s)
        \right\} 
        \right\rangle +\frac{\mu}{2} \left\langle x^2(t) \right\rangle .\notag
    \end{align}
The last term on the right hand side has already been calculated to yield Eq.~(\ref{sigma_xx}).
Using Eq.~(\ref{xsol}), we can write the first term denoted by $I_1(t)$ as
\begin{align}
I_1(t) &=  -\frac 1 2  \left\langle 
        \left\{
        x(t), B(t) 
        \right\} 
        \right\rangle \notag\\
        &=-\frac 1 {2m} \int_0^t ds\; G_2(s)D_1(s) ,
\end{align}
where $D_1(t)$ is given in Eq.~(\ref{D_1}).
The second term denoted by $I_2(t)$ can be calculated by using Eq.~(\ref{xsol}) as 
\begin{align}
  I_2(t)
    =& mF_{xx}(t) \langle x^2(0)\rangle + \frac 1 m F_{pp}(t) \langle p^2(0)\rangle \notag\\
    &+ F_{xp}(t) \frac 1 2 \langle \{x(0),p(0)\}\rangle + \widetilde{I}_2(t) ,
\end{align}
where
\begin{align}
& F_{xx}(t) = \int_0^t ds\;\eta(t-s) G_1(t)G_1(s), \label{Fxx}\\
& F_{pp}(t)=  \int_0^t ds\;\eta(t-s) G_2(t)G_2(s) ,\label{Fpp}
\end{align}
\begin{align}
 F_{xp}(t) =  \int_0^t ds\;\eta(t-s) &\Big( G_1(t)G_2(s) \notag \\
&+G_1(s)G_2(t)\Big), 
\label{Fxp}
\end{align}
and 
\begin{align}
 \widetilde{I}_2(t) =\frac 1 {2m}  &\int_0^t ds\;\eta(t-s) \int_0^t dt'\; \int_0^s dt'' \; G_2(t-t') \notag\\
  & \times G_2(s-t'') D_1(t'-t'').
\label{tildeI2}
\end{align}
In summary, we have
\begin{align}
    &\langle H_I(t)\rangle=  I_1(t)+I_2(t)+\frac{\mu}{2} \langle x^2(t)\rangle ,\\
     &\langle \Delta H_I \rangle = I_1(\tau)+I_2(\tau)+\frac{\mu}{2} \langle x^2(\tau)\rangle
     -\frac{\mu}{2} \langle x^2(0)\rangle.
\end{align}
Note that the change in the average interaction energy 
can be calculated when the initial state $\phi(0)=(H(0),L(0),C(0),1)$
is known by evaluating the above integrals over time.

For the Ohmic bath with the Lorentz-Drude cutoff considered in this paper with the spectral density $J(\omega)$ in 
Eq.~(\ref{J}), we have
\begin{align}
    \gamma(t) 
    =\frac{2}{m}\int_0^\infty d\omega\; \frac{J(\omega)}{\omega}\cos(\omega t)
    =2\gamma\Omega e^{-\Omega t} .
\end{align}
We have 
from Eq.~(\ref{etat}),
\begin{align}
    \eta(t)=\gamma'(t)=-2\gamma\Omega^2 e^{-\Omega t}
    \label{eta_t}
\end{align}
and, from Eq.~(\ref{mu}),
\begin{align}
    \mu=m\gamma(0) =2m\gamma\Omega .
\end{align}
We can now evaluate the integrals in Eqs.~(\ref{Fxx}), (\ref{Fpp}), 
(\ref{Fxp}) and (\ref{tildeI2}) using the expressions for Green's functions $G_1(t)$ and $G_2(t)$ in Eqs.~(\ref{G_2})
and (\ref{G_1}) and for the bath correlation function $D_1(t)$ in Eq.~(\ref{D_1:exp}).
The triple integral in Eq.~(\ref{tildeI2}) is quite involved, and we have used \texttt{MATHEMATICA} for symbolic integrations.

%

\end{document}